\documentclass{elsart}
\usepackage{color,amsmath,amsfonts,amssymb}

\usepackage[dvips]{graphics}
\usepackage{enumerate}

\usepackage{mark_newcommands}

\usepackage{mark_journals}

\newcommand{\boxedeqn}[2]{%
  \\ \\\fbox{
      \addtolength{\linewidth}{-2\fboxsep}%
      \addtolength{\linewidth}{-2\fboxrule}%
      \begin{minipage}{\linewidth}%
      \center {\bf #1}\\
      \begin{eqnarray}#2\end{eqnarray}%
      \end{minipage}%
    }\\ \\
}
\newcommand{\Psib}{{\bar \Psi}}
\newcommand{\deltab}{{\bar \delta}}

\newcommand{\remove}[1] {}

\begin{document} 
\begin{frontmatter}
\title{Implementation of the LANS-alpha turbulence model in a primitive equation ocean model}

\author[ccs]{Matthew W. Hecht}
\author[ccs,Imperial]{Darryl D. Holm}
\author[ccs,cnls]{Mark R. Petersen\corauthref{cor}}
\corauth[cor]{Corresponding author}
\ead{mpetersen@lanl.gov}
\author[ccs]{Beth A. Wingate}

\address[ccs]{Computer, Computational and Statistical Sciences Division,
 Los Alamos National Laboratory, Los Alamos, New Mexico}
\address[cnls]{Center for Nonlinear Studies,
 Los Alamos National Laboratory, Los Alamos, New Mexico}
\address[Imperial]{Mathematics Department, Imperial College  London, United Kingdom}

\begin{abstract}
This paper presents the first numerical implementation and tests of
the Lagrangian-averaged Navier-Stokes-alpha (LANS-$\alpha$) turbulence
model in a primitive equation ocean model. The ocean model in which we
work is the Los Alamos Parallel Ocean Program (POP); we refer to POP
and our implementation of LANS-$\alpha$ as POP-$\alpha$.

Two versions of POP-$\alpha$ are presented: the {\it full}
POP-$\alpha$ algorithm is derived from the LANS-$\alpha$ primitive
equations, but requires a nested iteration that makes it too slow for
practical simulations; a {\it reduced} POP-$\alpha$ algorithm is
proposed, which lacks the nested iteration and is two to three times
faster than the full algorithm.  The reduced algorithm does not follow
from a formal derivation of the LANS-$\alpha$ model equations.
Despite this, simulations of the reduced algorithm are nearly
identical to the full algorithm, as judged by globally averaged
temperature and kinetic energy, and snapshots of temperature and
velocity fields.  Both POP-$\alpha$ algorithms can run stably with
longer timesteps than standard POP. 

Comparison of implementations of full and reduced POP-$\alpha$
algorithms are made within an idealized test problem that captures
some aspects of the Antarctic Circumpolar Current, a problem in which
baroclinic instability is prominent. Both POP-$\alpha$ algorithms
produce statistics that resemble higher-resolution simulations of
standard POP.
   
A linear stability analysis shows that both the full and reduced
POP-$\alpha$ algorithms benefit from the way the LANS-$\alpha$
equations take into account the effects of the small scales on the
large. Both algorithms (1) are stable; (2) have an effective Rossby
deformation radius that is larger than the deformation radius of the
unmodeled equations; and (3) reduce the propagation speeds of the
modeled Rossby and gravity waves relative to the unmodeled waves at
high wave numbers.

\end{abstract}
\end{frontmatter}

\section{Introduction \label{s_introduction}}

Ocean-climate models are typically run at relatively low resolution
($1^\circ$ or $\sim100$km grid cells) in climate simulations due to the
computational requirements of the coupled components and duration of
the simulations, which might last for hundreds or thousands of model
years.  This resolution is well above the Rossby radius of deformation
over most of the ocean domain, the typical horizontal size of eddies in
the ocean.  As a result, ocean-climate simulations only include the
mean, large-scale flow, and not the eddies one might observe in
satellite images.  These eddies affect the mean circulation by
transporting buoyancy and kinetic energy.  Recent ocean-only
simulations at resolutions of $1/10^\circ$ and finer confirm that when
eddy-scale dynamics are resolved, some of the more prominent biases in
the mean circulation, such as the well known biases in Gulf Stream
path and structure, are greatly reduced \cite{Bryan_ea07om}.

The goal of turbulence modeling is to capture the effects of small
scale structures on the large-scale flow.  In the case of
ocean-climate models, this need is particularly pressing because the
Rossby radius, the length scale where available potential energy is
converted into kinetic energy, is not resolved except in the
equatorial region. This leaves not just {\it part} of the mesoscale
eddy spectrum unresolved, but {\it all} of that spectrum unresolved
except perhaps in the region in which the equatorial jets occur. Thus,
parameterization of those effects becomes a necessity.

A particular parameterization of scalar transport, following the
approach of Gent and McWilliams (GM) \cite{Gent_McWilliams90jpo} and
extensions thereof \cite{Griffies98jpo,Visbeck_ea97jpo} has been
widely embraced in ocean modeling. These schemes involve two sources
of scalar mixing. One of these is similar to horizontal mixing by
diffusion, but its diffusivity is rotated slightly from the horizontal
into a surface along which potential density is constant. The other
source of scalar mixing involves the diffusion of layer thickness,
which acts to flatten surfaces of constant density (isopycnal
surfaces). This flattening of density surfaces releases potential
energy and, thus, mimics the action of eddies generated through
baroclinic instability. (The scalars in question here include heat and
salt, the constituents of density.) This isopycnal mixing scheme,
first applied to coupled atmosphere-ocean models in the Community
Climate System \cite{Boville_Gent98jc}, represents one of the more
significant advances not only in ocean modeling, but more generally in
climate. Implementing this parameterization allows coupled atmosphere
and ocean models to be configured without the``flux corrections"
formerly required in order to compensate for incompatible heat
transports in oceanic and atmospheric components.

A number of important mean flow features in the oceans, however,
remain poorly represented in most ocean models. These flow features
only come into focus in eddy-resolving models which remain too
expensive for long time scale climate simulation. This resistance to
parameterization by scalar transport on isopycnal surfaces persists
in the Gulf Stream/North Atlantic Current system and in a few other
regions. This is probably because the process through which the
variability in the circulation feeds back on the mean is dynamically
more complex than what can be adequately parameterized by diffusion of
layer thickness. Hence, although the very serious issue of spurious
cross-frontal diffusion of heat (the ``Veronis effect"
\cite{Veronis75inbk,Boning_ea95jc}) is greatly reduced by using the GM
parameterization, the path and structure of the Gulf Stream and its
downstream development remain strongly biased.

In this paper we consider for the first time a non-diffusive
parameterization that primarily modifies the momentum equations. We
anticipate that this parameterization will prove complementary to the
scalar-transport-based parameterizations currently in use, a subject
for later investigation.

We have implemented a parameterization of mean multiscale transport
effects known as the Lagrangian-Averaged Navier-Stokes alpha
(LANS-$\alpha$) model in a primitive equation ocean model. The
LANS-$\alpha$ model is a member of a hierarchy of equations developed
using asymptotic methods and Lagrangian-averaging in Hamilton's
principle \cite{Holm99pd}.  These equations have desirable
characteristics, such as conservation of energy and potential
vorticity in the absence of dissipation. They also satisfy Kelvin's
circulation theorem and conserve a form of potential vorticity.

The LANS-$\alpha$ model is implemented by using two velocities: the
Lagrangian-averaged velocity and the Eulerian-averaged velocity.  The
Eulerian-averaged velocity is related to the Lagrangian-averaged
velocity through a smoothing operation. This smoother Eulerian
velocity, which does not include the smallest scale variability,
appears as the advecting velocity for both momentum and tracers.  The
Lagrangian-averaged (non-smoothed) velocity, which does include mean
effects of the smaller scales, is transported by this smoother
advecting velocity.  The parameter alpha is a length scale that
determines the amount of smoothing. Structures smaller than alpha are
slaved to the larger structures in the Eulerian-averaged velocity.
Solutions of the LANS-$\alpha$ equations converge to the Navier-Stokes
solutions in the limit as alpha goes to zero, and the LANS-$\alpha$
equations are Galilean invariant.  Note that the LANS-$\alpha$
equations change the {\it nonlinear} terms of the Navier-Stokes
equations, rather than the {\it dissipative} terms, as most turbulence
models, such as hyperviscosity, do.  In other words, LANS-$\alpha$ is
a model of stirring---the purely mechanical movement and rearrangement
of fluid---rather than mixing, which simply occurs through diffusion
(see, e.g. \cite{Aref99n}).  The LANS-$\alpha$ model also tempers vortex
stretching while preserving the original form of the vortex dynamics
\cite{Foias_ea01pd}.  This can be seen with the vorticity equation,
where the vortex stretching term contains the smoothed velocity.  The
LANS-$\alpha$ equations may be interpreted as a closure of a viscous
version of the GLM equations of Andrews and
McIntyre\cite{Andrews_McIntyre78jfm}.

The alpha-parameterization was introduced within inviscid equations by
Holm, Marsden and Ratiu \cite{Holm_ea98prl}. Soon afterwards, its
viscous version, LANS-alpha, was developed into a turbulence model by
Chen et al. \cite{Chen_ea99pd_a,Chen_ea99pd_b}. See also Marsden and
Shkoller \cite{Marsden_Shkoller01ptrsa} for parallel developments. Its
performance as a model of turbulence and mixing was tested in a
variety of observations and numerical simulations.  A summary of
results obtained between 1997 and 2004 can be found in
\cite{Holm_ea05las}; only the highlights are reviewed here.  In
high-Reynolds-number pipe flow, its steady analytical solutions
matched observations of mean velocity over several orders of
magnitude, demonstrating its potential for describing the mean
properties of turbulence over a wide range of scales
\cite{Chen_ea99pd_a}. In forced isotropic turbulence in a
three-dimensional periodic domain, the energy spectrum $E(k)$ was
found to be proportional to the expected $k^{-5/3}$ for $k\alpha<1$
\cite{Chen_ea99pd_b}.  However, for $k\alpha>1$ the energy spectrum
turns over to approximate $k^{-3}$, thereby greatly reducing the
number of degrees of freedom excited in the flow. This occurs because
the scales smaller than $\alpha$ (that is, $k\alpha>1$) are {\it
slaved} to the larger scales. This slaving greatly reduces the
computational work required in a numerical fluid simulation.  In fact,
the computational work for a turbulence simulation scales with the
Reynolds number $Re$ as $Re^3$ for Navier-Stokes and as $Re^2$
LANS-$\alpha$. The LANS-$\alpha$ model was found to be at least as
accurate as modern Large Eddy Simulation (LES) models (such as the
dynamic models in \cite{Germano_ea91pf,Meneveau_Joseph00arfm}) when
measured against a reference solution in a high-resolution direct
numerical simulation of Kelvin-Helmholtz shear layer instability
\cite{Geurts_Holm02inbk}.

Recent work has extended LANS-$\alpha$ to geophysical fluid dynamic
models.  For example, a quasi-geostrophic alpha model, developed by
Holm and Nadiga \cite{Holm_Nadiga03jpo}, showed that the alpha model
can capture the qualitatively correct time-mean circulation in the
double-gyre problem at a resolution four to eight times coarser than a
traditional quasi-geostrophic model.  The shallow water LANS-$\alpha$
equations have been shown to allow larger time steps than the
unaveraged shallow water equations \cite{Wingate04mwr}.  A linear
analysis leads to the understanding that this is due to a slow down of
the frequency of the high wave-number gravity waves.  As the grid is
refined, the time step becomes independent of the mesh spacing and
dependent on $\alpha$.  The LANS-alpha model's method of accounting
for the effect of the small scales on the large impacts the way it
models baroclinic instability in a two-layer QG model; it lowers the
critical wavenumber, reduces the bandwidth of the instability, and
preserves the value of forcing at the onset \cite{Holm_Wingate05jpo}.

Our work presents the first implementation of the LANS-$\alpha$ model
in a primitive equation ocean general circulation model, and therefore
the first that could potentially be used for century to millennium
scale climate projection.  Implementation of LANS-$\alpha$ in Los
Alamos' Parallel Ocean Program (POP) presented numerous new
challenges, including most notably accommodating the
barotropic/baroclinic mode splitting that is performed in ocean
models of this class. This decomposition of the momenta into a
depth-averaged barotropic component and the three dimensional
baroclinic residual is responsible for a gain of efficiency on the
order of one to two orders of magnitude under explicit or
semi-implicit time-stepping, and so must be preserved, despite the
complication presented to implementation of LANS-$\alpha$. Other
challenges involved enforcement of continuity on the Eulerian-averaged
velocity in the barotropic equations; correct handling of boundaries
and topography; and establishment of an efficient algorithm that
retains the essential properties of the LANS-$\alpha$ model.

The organization of this paper is as follows: after reviewing the
LANS-$\alpha$ model in \S2, the POP-$\alpha$ algorithm is derived from
the LANS-$\alpha$ primitive equation set in \S3.  A reduced version of
the full POP-$\alpha$ algorithm, which is faster by a factor of two to
three, is presented in \S3.5.  The stability analysis in \S4 shows
that the effect of the LANS-$\alpha$ model is to increase the
effective Rossby radius of deformation, lowering the wavenumber at
which the onset of baroclinic instability occurs, and to slow down
gravity waves and Rossby waves at high wave numbers. In this section
we also show that the reduced algorithm does not change the character
of the continuous equations very much, and that both the full and
reduced algorithms are stable.  Simulations using an idealized channel
domain that invokes baroclinic instability are presented in \S5.
They show that POP-$\alpha$ statistics resemble higher-resolution
simulations of standard POP, and that results from the full and
reduced POP-$\alpha$ algorithms are nearly identical. Section \S6
states our conclusions.

\section{LANS-$\alpha$ model equations}
\subsection{The LANS-$\alpha$ model}

The nondissipative part of the LANS-$\alpha$ equations is obtained
using the Euler-Poincar{\'e} variational framework, where the
Lagrangian expression of Hamilton's principle is obtained for each
approximation of the full equations of motion, such as shallow water,
quasi-geostrophic, Boussinesq, and primitive equations
\cite{Holm99pd,Holm_ea98inbk}. The effect of this procedure is that
two fluid velocities appear: the Lagrangian-averaged velocity, $\bv$,
which is the momentum velocity, and the Eulerian-averaged velocity,
$\bu$, which is the advecting velocity.  In the original derivation,
the Lagrangian-averaged and Eulerian-averaged velocities are related
by the Helmholtz operator, \bea \bv = \left( 1- {\tilde \Delta}
\right) \bu, \;\;\; {\tilde \Delta} \equiv \del \cdot\left<{\bf
\xi}{\bf \xi}\right>\cdot \del
\label{Helmholtz1}
\eea 
where the displacement fluctuation $\xi={\bf X}-\left<{\bf X}\right>$
is the difference between the spatial trajectory of a fluid parcel,
${\bf X}$, and the Lagrangian mean of that fluid trajectory,
$\left<{\bf X}\right>$.  The covariance of the displacement
fluctuation, $\left<{\bf \xi}{\bf \xi}\right>$, is a three-by-three
symmetric tensor (in three-dimensional turbulence) that describes the
small-scale (or short-time) fluctuations in the fluid flow.  This
tensor varies in both space and time.  The size of the fluctuations
described by $\left<{\bf \xi}{\bf \xi}\right>$ is determined by the
time-scale of the Lagrangian mean time-averaging operation
$\left<\,\cdot\,\right>$, as described by Holm \cite{Holm99pd},
p. 217. It is possible to replace (\ref{Helmholtz1}) by a convolution
(and thus, a filter), as discussed in Geurts and Holm
\cite{Geurts_Holm03pf}.

The Lagrangian mean equations include a prognostic equation for each
of the components of $\left<{\bf \xi}{\bf \xi}\right>$, which are then
used in (\ref{Helmholtz1}) and the Navier-Stokes equation.  The
additional memory and computation required by this fully dynamic
covariance tensor could be prohibitively expensive for computational
fluids models.  In the original LANS-$\alpha$ model, the covariance
displacement fluctuation tensor $\left<{\bf \xi}{\bf \xi}\right>$ is
simplified to the scalar $\alpha^2$ by assuming that the displacements
are spatially isotropic, uniform throughout the domain, and constant
in time.  These are severe restrictions to the original Lagrangian
mean equations.  Nevertheless, the resulting LANS-$\alpha$ equation
set has been shown to be an effective turbulence model
\cite{Chen_ea99pd_a,Holm_Nadiga03jpo}.

\subsection{The primitive equation LANS-$\alpha$ model}

The implementation of the LANS-$\alpha$ model in POP described in this work uses the primitive-equation form (\cite{Holm_ea98inbk}, section 6.3):
\bea
&\dd{\bv}{t} + \bu\cdot\del\bv + u_3\p_z\bv
   + v_j \del u_j 
   + {\bf f}\times \bu
 = -\frac{1}{\rho_0}\del \pi 
   + {\mathcal F}\left({\bv}\right) 
\label{PE1}\\
& \dd{\pi}{z}=-\rho g, \label{hydrostatic1} \\
&\del\cdot\bu + \p_z u_3 = 0\label{cont1}, \\
&\bv = \left(1-\alpha^2(\del^2 +\p_z^2)\right)\bu,
\label{Helm1} 
\eea
where $\bv=(v_1,v_2)$ and $\bu=(u_1,u_2)$ are horizontal velocities
and $\del$ is the horizontal gradient.  The Helmholtz operator in
(\ref{Helm1}) indicates that the Eulerian-averaged velocity $\bu$ is
smoother than the Lagrangian-averaged velocity $\bv$.  In the
Lagrangian mean equations, the covariance displacement fluctuation
tensor $\left<{\bf \xi}{\bf \xi}\right>$ is determined prognostically
by the equations.  In the LANS-$\alpha$ model, alpha takes the place
of this tensor; this length scale alpha is the only user-defined
parameter in the model.  Alpha determines how smooth $\bu$ is; more
precisely, structures in $\bv$ which are smaller than alpha are
strongly suppressed by the inverse Helmholtz operator and do not
appear in $\bu$.  In practice alpha is chosen to be on the order of a
grid-cell width.  An important question is how well the effects of
turbulence are represented in the particular context of a primitive
equation ocean model as alpha is varied; this is addressed in
\cite{Petersen07jcp}.

A fundamental characteristic of the LANS-$\alpha$ equations is that
the smoother Eulerian-averaged velocity $\bu$ is the advecting
velocity.  Thus the small-scale structures in $\bv$ are advected by
the smoother velocity $\bu$. However, this advection is not like
passive tracers embedded in the flow. Rather, it is fluid transport as
in Kelvin's circulation theorem, where the line element with respect
to which the circulation is defined must also be advected. This is the
source of the additional term $\sum_{j=1}^{3} v_j \del u_j =
(\del\bu)^T\cdot\bv$ in the motion equation for the alpha model.

The diffusion operator ${\mathcal F}$ acts on the Lagrangian-averaged
velocity $\bv$, so (\ref{PE1}) can be thought of as an
advection-diffusion equation for $\bv$, and $\bv$ as the specific
momentum.  Additional terms are the Coriolis force ${\bf f}\times
\bu$, the pressure gradient ${\rho_0}^{-1}\del \pi$, and the
additional nonlinear term, $\sum_{j=1}^3 v_j \del u_j$.  In the
absence of this extra nonlinear term, the LANS-$\alpha$ model reduces
to the Leray model \cite{Holm99pd}.  A comparison of the LANS-$\alpha$
and Leray models will be presented in a future work.

The pressure $\pi$ is not the physical pressure but a modified pressure, where
\bea
\pi = p -\frac{1}{2}|\bu|^2 - \frac{\alpha^2}{2}|\del \bu|^2
\eea
and $p$ is the actual pressure.  However, in the algorithms presented in this paper, the modified pressure $\pi$ is never calculated this way.  Instead, it is computed at each time step from the hydrostatic equation (\ref{hydrostatic1}) once the density at that step is known.  This is identical to the way the physical pressure is computed in POP.

The LANS-$\alpha$ version of tracer equations in a primitive-equation ocean model is
\bea
&\dd{\varphi}{t} + \bu\cdot\del\varphi 
 = {\mathcal D}\left({\varphi}\right),
\eea
where $\varphi$ may be temperature, salinity, or another conserved quantity, and ${\mathcal D}$ is the horizontal and vertical tracer diffusion.  The tracer advecting velocity is the smooth velocity $\bu$.  This is the velocity responsible for transport of fluid particles carrying mass and salinity. 
This is not a model choice; rather it is part of the derivation of the alpha model. Before their diffusivities were introduced as kinetic coefficients, the tracer equations described quantities such as heat and salinity that were simply carried along with each fluid parcel. The Lagrangian average of such a quantity that is conserved on fluid parcels is simply its average value, computed at fixed Lagrangian coordinate. Likewise, the Lagrangian averaged equation for such a process keeps the same form as the exact dynamics, but the transport velocity in this equation is replaced by the Lagrangian averaged particle velocity ($\bu$) and the instantaneous value of the tracer is replaced by its average value taken along the trajectory of the fluid parcel. The value of the diffusivity of such a Lagrangian averaged tracer is a model parameter. However, the form of its dynamics was fixed by the use of the Lagrangian average in the derivation of the alpha model.  

\section{Implementation of LANS-$\alpha$ in POP}
\subsection{POP barotropic/baroclinic splitting}
The Parallel Ocean Program (POP), developed at Los Alamos National Laboratory by Smith, Dukowicz and Malone \cite{Smith_ea92pd}, is well-known in the ocean modeling community.  As the ocean component of NCAR's Community Climate System Model \cite{Collins_ea06jc} and NCAR's Parallel Climate Model, it is used for twenty-first century climate simulations by the Intergovernmental Panel on Climate Change (IPCC).

POP uses finite differences to discretize the primitive equations, and has a leap-frog time stepping scheme.  The horizontal grid is 
a logically rectangular B-grid, the vertical coordinate is z-level and incorporates a free surface \cite{Dukowicz_Smith94jgro} and optional partial bottom cells \cite{POP_manual}. 
Diffusion can be Laplacian or biharmonic, and the Gent-McWilliams scheme is available for tracer diffusion \cite{Gent_McWilliams90jpo}.

The barotropic/baroclinic mode splitting used in ocean general circulation models presents a significant complication to the implementation of LANS-$\alpha$. The mode splitting is performed in order to isolate the very fast free-surface gravity waves, which have speeds in excess of 200 m/sec. These fast external waves are responsible for the very rapid propagation of tsunami events, for example,   but have negligible bearing on ocean circulation. The problematic external mode is almost completely isolated by a modal decomposition into the vertically averaged velocity and the departure from that average, with internal waves speeds of a few meters per second, at most, presenting a much less stringent time step limitation within the 3-dimensional baroclinic equations. The much faster external mode is treated by solving the barotropic equations with an implicit method in POP; some other models use an explicit subcycling of the barotropic equations. In either case the isolation of the external mode within a lower-dimensional set of equations results in a tremendous gain in efficiency. It is within these split equations that the model is implemented in order to arrive at a primitive equation ocean model containing  the LANS-$\alpha$ turbulence scheme in a form suitable not just for the very limited research application that could be addressed with a simpler unsplit model, but for consideration in climate modeling.


In POP, the barotropic velocity $\bU(x,y,t)=(U_1,U_2)$ is defined as the vertical integral of the horizontal velocity $\bu(x,y,z,t)=(u_1,u_2)$,
\bea
\bU \equiv \frac{1}{H+\eta}\int_{-H}^\eta \bu \;\;dz,
\eea
where $H(x,y)$ is the ocean depth when the surface is at rest and $\eta(x,y,t)$ is the free surface height.  Subscripts are used on $U$ and $u$ to indicate horizontal directions because 
we reserve 
$V$ and $v$ 
for the rough velocity in the LANS-$\alpha$ model.

The continuity equation may now be integrated in the vertical to produce a prognostic equation for the free surface height,
\bea
&\int_{-H}^\eta \left( \del\cdot\bu + \p_zu_3\right) dz=0,\\
&\dd{\eta}{t} + \del \cdot \left(H+\eta\right)\bU = 0.\label{eta1}
\eea

\begin{figure}[tbh]
\center
\scalebox{.65}{\includegraphics{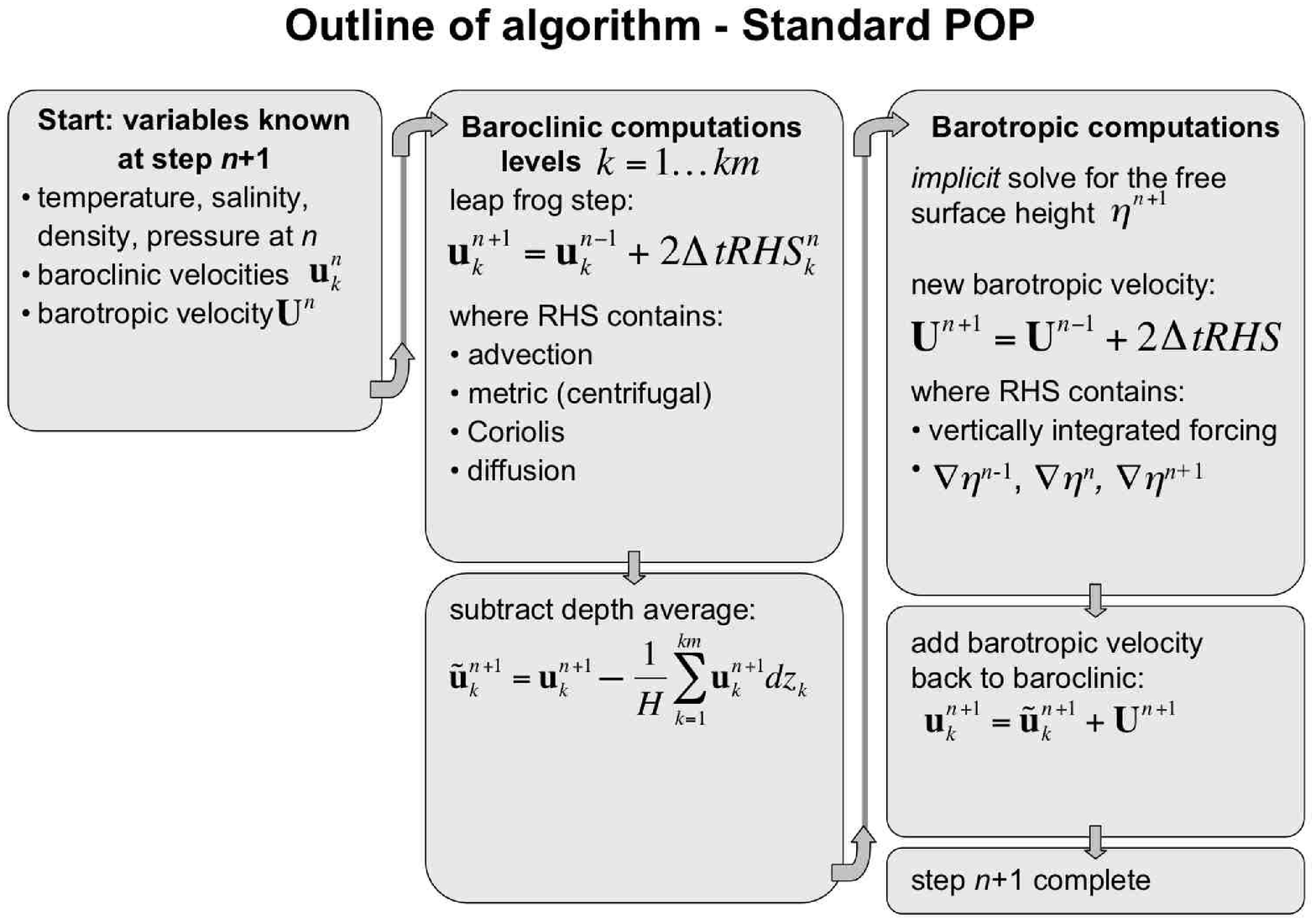}} 
\caption{\label{f_POP_outline} 
Schematic of POP algorithm, showing the baroclinic/barotropic splitting.
}\end{figure}

An outline of the POP algorithm for the momentum equation using the baroclinic/barotropic splitting at time step $n+1$ is presented graphically in Fig. \ref{f_POP_outline} and as follows:
\begin{enumerate}[({P}1)]
\item Baroclinic component\\
leapfrog step: $\bu_k^{n+1} = \bu_k^n + 2\Delta t RHS^n_k$,\;\;\; $k=1\ldots km$\\
where $RHS^n_k$ contains the momentum forcing: advection, centrifugal, Coriolis, diffusion, and pressure gradients.
\item Subtract depth-average:\\
${\tilde \bu}_k^{n+1} = \bu_k^{n+1} - \frac{1}{H}\sum_{\kappa=1}^{km}\bu_\kappa^{n+1}dz_\kappa, 
 \;\;\;k=1\ldots km$
\item Barotropic component:\\
compute $\eta^{n+1}$ using the conjugate gradient method.\\
leapfrog step: $\bU^{n+1} = \bU^{n-1} + RHS^n$\\
where $RHS^n$ contains $\del \eta$, 
the Coriolis term, and the vertically integrated forcing.
\item Add baroclinic and barotropic velocities: \\
$\bu_k^{n+1} = {\tilde \bu}_k^{n+1} + \bU^{n+1},\;\;\; k=1\ldots km$
\end{enumerate}

In this notation, $\bu_k^n$ is the baroclinic horizontal velocity at the $k$th vertical level at time step $n$, for levels $k=1\ldots km$, $\Delta t$ is the time step, and ${\tilde \bu}$ is an intermediate baroclinic velocity.  This outline contains only enough detail for the purposes of this paper; a complete description can be found in the POP reference manual \cite{POP_manual}.

\subsection{POP-$\alpha$ baroclinic implementation}
Because POP's baroclinic and barotropic components use explicit and implicit time stepping, respectively, the implementation of the alpha model in POP was quite different in these two parts of the code.  The explicit baroclinic part was straightforward, while the implicit barotropic part presented numerous difficulties, as detailed in the following section.

When POP is adapted to use the LANS-$\alpha$ model (\ref{PE1}--\ref{Helm1}), the following changes must be made: (1) there are two full velocity fields, the Eulerian-averaged, or smooth, velocity $\bu$ and the Lagrangian-averaged, or rough, velocity $\bv$; (2) likewise, there is a smooth and rough barotropic velocity, $\bU$ and $\bV$; (3) the appropriate velocities must be used in each term in the momentum equation; (4) the nonlinear term $\del \bu^T\cdot\bv$ is added to the forcing terms in the momentum equation; and (5) the advecting velocity in the tracer equation is the smooth velocity $\bu$.

Because the time derivative in the LANS-$\alpha$ momentum equation (\ref{PE1}) is on the rough velocity $\bv$, one must take a time step in $\bv$ and then compute $\bu$ by inverting the Helmholtz operator (\ref{Helm1}), i.e.
\bea
\frac{\bv^{n+1}_k-\bv^{n+1}_k}{2\Delta t} 
  + \bu^n_k\cdot\del\bv^n_k + u^n_{3,k}\p_z\bv^n_k 
  + \sum_{j=1}^2 u^n_{j,k}\del v^n_{j,k} 
   + {\bf f}\times \bu^n_k \nn \\ 
 = -\frac{1}{\rho_0}\del p^n_k + {\mathcal F}{\bv^n_k},
\label{PE2}\\
\bu^{n+1}_k = \left(1-\alpha^2\del^2\right)^{-1}\bu^{n+1}_k
,\;\;\; k=1\ldots km
\label{Helm2} 
\eea
where $\bv^{n}_k$ and $\bu^{n}_k$  are the rough and smooth horizontal velocities at time step $n$ and vertical level $k$, $v^n_{j,k}$ is the $j^{th}$ component at level $k$, and $\del$ is the horizontal gradient.

\begin{figure}[tbh]
\center
\scalebox{.65}{\includegraphics{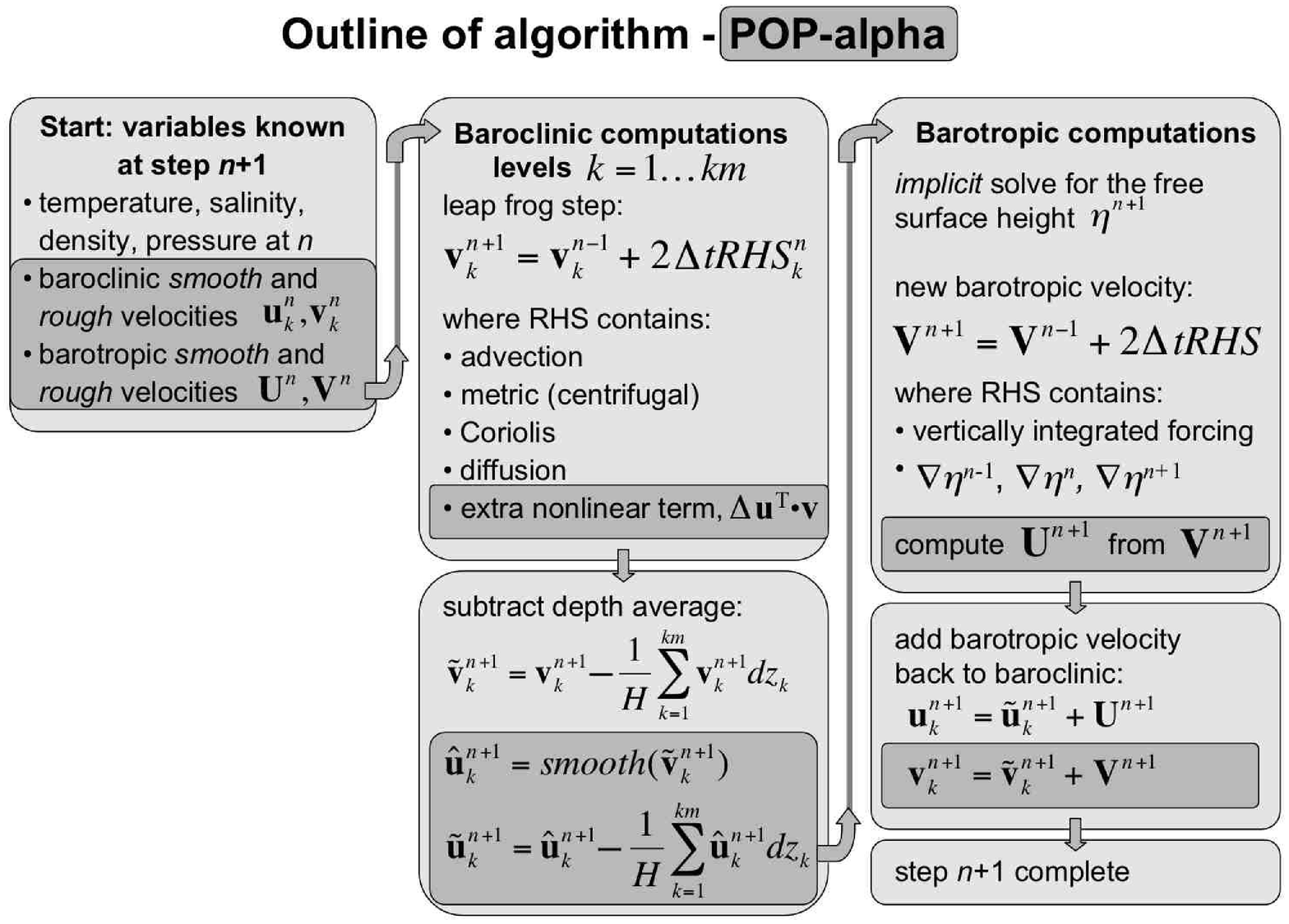}} 
\caption{\label{f_POPa_outline} 
Schematic of POP-$\alpha$ algorithm, showing the baroclinic/barotropic splitting.
}\end{figure}

The outline for the POP-$\alpha$ algorithm is then
\begin{enumerate}[({A}1)]
\item Baroclinic component\\
leapfrog step: $\bv_k^{n+1} = \bv_k^n + 2\Delta t RHS^n_k$,\;\;\; $k=1\ldots km$\\
where $RHS^n_k$ contains the momentum forcing:
 \framebox{extra nonlinear term $ u^n_{j,k}\del v^n_{j,k} $},
 advection, centrifugal, Coriolis, diffusion, and pressure gradients.
\item Subtract depth-average from $\bv$:\\
${\tilde \bv}_k^{n+1} = \bv_k^{n+1} - \frac{1}{H}\sum_{\kappa=1}^{km}\bv_\kappa^{n+1}dz_\kappa, 
 \;\;\;k=1\ldots km$
\item Solve for smooth baroclinic velocity: \\
\framebox{$\buh^{n+1}_k = \left(1-\alpha^2\del^2\right)^{-1}
   {\tilde \bv}^{n+1}_k,\;\;\; k=1\ldots km$}
\item  Subtract depth-average from $\buh$:\\
\framebox{${\tilde \bu}_k^{n+1} = \buh_k^{n+1} 
  - \frac{1}{H}\sum_{\kappa=1}^{km}\buh_\kappa^{n+1}dz_\kappa, 
 \;\;\;k=1\ldots km$}
\item Barotropic component:\\
compute $\eta^{n+1}$ using the conjugate gradient method.\\
leapfrog step: $\bV^{n+1} = \bV^{n-1} + RHS^n$\\
where $RHS^n$ contains $\del \eta$, Coriolis term, and the vertically integrated forcing.
\item \framebox{solve for smooth barotropic velocity, $\bU^{n+1}$ }
\item Add baroclinic and barotropic velocities: \\
$\bv_k^{n+1} = {\tilde \bv}_k^{n+1} + \bV^{n+1},\;\;\; k=1\ldots km$\\
\framebox{${\tilde \bu}_k^{n+1} = \bu_k^{n+1} + \bU^{n+1},\;\;\; k=1\ldots km$}
\end{enumerate}
The boxes indicate steps which were added to POP to implement the LANS-$\alpha$ model.  Velocities ${\tilde \bv}, {\tilde \bu}$, and $\buh$ are intermediate baroclinic velocities.  Fig. \ref{f_POPa_outline} shows a graphical version of the POP-$\alpha$ algorithm.  The depth-average must to be subtracted from $\bv$ and then again from $\bu$ in steps (A2) and (A4) to guarantee that both have zero barotropic component.   

Note that in the Helmholtz operator
$\left(1-\alpha^2\del^2\right)^{-1}$, the smoothing is in the
horizontal directions only.  This is a reduction from the original
LANS-$\alpha$ formulation (\ref{Helm1}), which includes a vertical
component.  This reduction was chosen not only to reduce computational
time, but because it is numerical resolution in the horizontal plane
which limits physical processes at the scale of the Rossby radius of
deformation and presents the opportunity for effective use of the
turbulence model.

The extra nonlinear term---the fourth term in (\ref{PE2})---only includes a summation from $j=1\ldots2$ rather than $j=1\ldots3$ as in the original LANS-$\alpha$ equations.  The missing term, $u_3\del v_3$, involves only vertical velocities and is several orders of magnitude smaller that the first two terms in the summation in primitive equation applications.  If this term were included a smoothed vertical velocity $u_3$ would have to be computed at each horizontal level, adding to the overall computational cost for this unimportant term.

Simulations where LANS-$\alpha$ was implemented in only the baroclinic component (A1--A4) were immediately unstable, failing to converge within 100 timesteps.  Those with LANS-$\alpha$ only in the barotropic component (A5--A6) of the algorithm were usually unstable, and failed to converge after $10^5$ timesteps (30 years).  Thus LANS-$\alpha$ must be uniformly implemented in both components.

\subsection{POP barotropic implementation}
The abbreviated description of the barotropic component in steps 5 and 6 on the POP-$\alpha$ algorithm above does not reveal the full complexity of these steps.  Essentially, one needs to find $\bV^{n+1}$,  $\bU^{n+1}$, and $\eta^{n+1}$ from the vertically integrated momentum and continuity equations by the end of the procedure.  However, both of these equations depend on the velocities and surface height, so there are potentially two simultaneous implicit solves.  

To review this difficulty in more detail, we step back from the LANS-$\alpha$ model and review the POP barotropic implementation by Dukowicz and Smith \cite{Dukowicz_Smith94jgro}.  The leapfrog discretization of the barotropic momentum equation and vertically integrated continuity equation are:
\bea
&\ds \frac{\bU^{n+1}-\bU^{n-1}}{2\Delta t} +  {\sf B}\bU^{\xi'}
  = - g\del\eta^\xi +  {\bf G}^n, \label{mom_stPOP} \\
&\ds \frac{ \eta^{n+1}-\eta^n}{\Delta t}
   + \del\cdot H\bU^\theta  =0, \label{cont_stPOP2} 
\eea
where $g$ is gravitational acceleration, ${\bf G}^n=(G_x^n,G_y^n)$ are the $x$ and $y$ components of the vertically integrated baroclinic momentum forcing terms, and the Coriolis parameter is written as the matrix
\bea
{\sf B} = \left[  \ba{cc} 0& -f\\ f & 0\ea   \right].
\eea
The superscripts $\zeta$, $\zeta'$, and $\theta$ are parameters which specify the degree to which the Coriolis term, pressure gradient, and continuity are implicit:
\bea
& \bU^{\xi'} 
  = \xi'\bU^{n+1} + (1-\xi' - \gamma')\bU^n + \gamma'\bU^{n-1},\label{def_U_xi}\\
& \eta^{\xi} 
  = \xi\eta^{n+1} + (1-\xi - \gamma)\eta^n + \gamma\eta^{n-1},\label{def_eta_xi}\\
& \bU^{\theta} = \theta \bU^{n+1} + (1-\theta)\bU^n. \label{def_U_theta}
\eea
In the POP code, the pressure splitting is equally weighted ($\xi=\gamma=1/3$) and the divergence is fully implicit ($\theta=1$).  These choices are the optimal choice of parameters to damp the computational gravity wave mode (\cite{Dukowicz_Smith94jgro}, Fig. 2).  One may specify implicit Coriolis ($\xi'=\gamma'=1/3$) or centered explicit Coriolis ($\xi'=\gamma'=0$) in the POP input file.  The advantage of an implicit Coriolis term is that a larger time step may be taken, sometimes by a factor of two.  The POP-$\alpha$ algorithm described in this paper uses an explicit Coriolis term because the implicit version would require an iterative method to solve the barotropic momentum equation.

Using $\xi=\gamma=1/3$, $\theta=1$, and explicit Coriolis ($\xi'=\gamma'=0$), equations (\ref{mom_stPOP}---\ref{cont_stPOP2}) become
\bea
&\bU^{n+1} = \bU^{n-1} + \tau \left[ {\bf G}^n - {\sf B}\bU^n
   -g\gamma\del\left(\eta^{n+1} + \eta^n + \eta^{n-1} \right)
     \right], \label{momx_stPOP_expC1} \\
&\frac{2}{\tau}\left( \eta^{n+1}-\eta^n \right) 
   + \del\cdot H\bU^{n+1}  =0, \label{cont_stPOP_expC1} 
\eea
where $\tau=2\Delta t$.  These equations are both implicit and so must be solved simultaneously for $\bU^{n+1}$ and $\eta^{n+1}$.  Substituting $\bU^{n+1}$ from (\ref{momx_stPOP_expC1}) into (\ref{cont_stPOP_expC1}) and solving for the pressure, we have
\bea
&&   \left( \del\cdot H\del -\frac{2}{\gamma g\tau^2} \right)\eta^{n+1}
 =   -\frac{2}{\gamma g\tau^2} \eta^n
   + \frac{1}{\gamma \tau g}\nabla\cdot H\bU^{n-1} 
   \nn\\&& \;\;\;\;\;\;
   + \frac{1}{g\gamma}\del\cdot H \left[ {\bf G}^n - {\sf B}\bU^n
   -g\gamma\del\left( \eta^n + \eta^{n-1} \right),
   \right] 
 \label{cont_stPOP_expC3} \\
&&\bU^{n+1} = \bU^{n-1} + \tau \left[ {\bf G}^n - {\sf B}\bU^n
   -g\gamma\del\left(\eta^{n+1} + \eta^n + \eta^{n-1} \right)
     \right]. \label{momx_stPOP_expC1b} 
\eea
Thus the algorithm for solving the barotropic equations with explicit Coriolis terms is to solve (\ref{cont_stPOP_expC3}) for $\eta^{n+1}$ and then (\ref{momx_stPOP_expC1b}) for $\bU^{n+1}$.  Solving for $\eta^{n+1}$ requires the inversion of a Laplacian-like operator; POP uses a conjugate gradient (CG) routine.

\subsection{Full POP-$\alpha$ barotropic implementation\label{s_full_alpha}}
The POP-$\alpha$ version of the barotropic equations (\ref{mom_stPOP}-\ref{cont_stPOP2}) are:
\bea
&\bV^{n+1}-\bV^{n-1} + \tau {\sf B}\bU^{\xi'}
  = -\tau g\del\eta^\xi + \tau {\bf G}^n, \label{mom_POPa} \\
&\frac{2}{\tau}\left( \eta^{n+1}-\eta^n \right) 
   + \del\cdot H\bU^\theta  =0, \label{cont_POPa} \\
&\bV^{n+1} = \left(1-\alpha^2\del^2\right)\bU^{n+1}. \label{Helm_btrop}
\eea
Using the same weighting choices as POP with explicit Coriolis terms, we have 
\bea
& \bV^{n+1} - \bV^{n-1}
   = \tau \left[{\bf G}^n -  {\sf B}\bU^n
   - \gamma g\del\left(\eta^{n+1} + \eta^n + \eta^{n-1} \right)
     \right], \label{mom_POPa_expC} \\
&\frac{2}{\tau}\left( \eta^{n+1}-\eta^n \right) 
   + \del\cdot H\bU^{n+1}  =0. \label{cont_POPa_expC} 
\eea
To solve equations (\ref{Helm_btrop}---\ref{cont_POPa_expC}) for $\bV^{n+1}$, $\bU^{n+1}$, and $\eta^{n+1}$, replace $\bV^{n+1}$ and $\bV^{n-1}$ in (\ref{mom_POPa_expC}) with $\bU^{n+1}$ and $\bU^{n-1}$  using the Helmholtz relation, solve for $\bU^{n+1}$, substitute into (\ref{cont_POPa_expC}), and solve for $\eta^{n+1}$.  The algorithm is then
\boxedeqn{Full POP-$\alpha$ barotropic algorithm:}{
&&   \left( \del\cdot H \left(1-\alpha^2\del^2\right)^{-1}\del
             -\frac{2}{\gamma g\tau^2} \right)\eta^{n+1}
 =   -\frac{2}{\gamma g\tau^2} \eta^n
   + \frac{1}{\tau\gamma g}\nabla\cdot H\bU^{n-1} 
\nn\\&& \;\;\;\;\;\;
   + \frac{1}{g\gamma} \del \cdot H 
       \left(1-\alpha^2\del^2\right)^{-1}\left[
   {\bf G}^n - {\sf B}\bU^n
   -g\gamma\del\left( \eta^n + \eta^{n-1} \right) 
   \right], 
   \label{full_eta} \\
&& \bV^{n+1} =  \bV^{n-1}
     + \tau \left[ {\bf G}^n -{\sf B}\bU^n
   - \gamma g\del\left(\eta^{n+1} + \eta^n + \eta^{n-1} \right)
    \right], 
    \label{full_V} \\
&&\bU^{n+1} = \left(1-\alpha^2\del^2\right)^{-1}\bV^{n+1}. 
\label{full_U}
}

The added complexity of the algorithm due to the LANS-$\alpha$ model in these equations is the smoothing step wherever the $\left(1-\alpha^2\del^2\right)^{-1}$ operator appears.  This Helmholtz inversion is solved using the iterative CG method.  The operator in front of $\eta^{n+1}$ in (\ref{full_eta}) is also inverted using the CG method.  This means that solving the full algorithm requires solving a {\it nested} inversion.  There are typically 50 iterations of CG in standard POP, so here there might be $50^2$ iterations to solve for $\eta^{n+1}$ when the Helmholtz inversion is used.  

Another option is to use a simple filter such as nearest-neighbor
averaging instead of a Helmholtz inversion to avoid the nested
iteration.  Filters have been successfully used in LANS-$\alpha$
models in large eddy simulations by Geurts and Holm
\cite{Geurts_Holm05jt}.  Using a local filter in place of the
Helmholtz inversion improves speed, but the filter must still be
applied within each CG iteration for the $\eta^{n+1}$ solve.

The best option would be to avoid the smoothing operator on the LHS of (\ref{full_eta}) altogether.  This is the motivation for the reduced algorithm presented in the next section.  Data on the computing time for these different options are discussed in section \ref{ss_results_compare}

Another version of the algorithm (\ref{full_eta}---\ref{full_U}) would be to replace (\ref{full_U}) with
\bea
&& \bU^{n+1} = \bU^{n-1} \nn\\&&\;\;\;
    + \tau \left(1-\alpha^2\del^2\right)^{-1} \left[ 
   {\bf G}^n - {\sf B}\bU^n
   -g\gamma\del\left(\eta^{n+1} + \eta^n + \eta^{n-1} \right) 
   \right],
 \label{full_U_alt} 
\eea
which can be obtained by combining (\ref{full_V}), (\ref{full_U}), and
(\ref{Helm_btrop}).  Although this is a valid derivation of the model
equations, it was found to be unstable in practice.  In
(\ref{full_U_alt}), $\bU^{n+1}$ is calculated from $\bU^{n-1}$ using a
pressure-averaged leap-frog time step.  In practice this allows the
smooth velocity $\bU$ to drift away from the rough velocity $\bV$:
Numerical experiments where this algorithm is used proceed as expected
initially, but after five or ten years (order $10^6$ time steps) $\bU$
is seen to depart greatly from $\bV$.  This drift is avoided by using
(\ref{full_U}), with $\bU$ computed as a smooth version of $\bV$ at
every step.

There are two types of variables that are smoothed in the full POP-$\alpha$ algorithm: the velocity in (\ref{full_U}) and the pressure gradients in (\ref{full_eta}).  The form of the pressure gradient term is $\left(1-\alpha^2\del^2\right)^{-1}\del\eta$, 
which comes directly from the derivation of the equations and was used in the algorithm.  This form requires boundary conditions for the pressure gradient on the boundary, which are zero.  In a periodic domain, the order of the filter and gradient could be changed, that is, 
$\del\left(1-\alpha^2\del^2\right)^{-1}\eta$
 could be used instead.  This change in the operator order is not possible with solid boundaries, as then the free surface height $\eta$ must be specified at the boundary, and it is unknown.  Thus the Helmholtz inversion must be applied to the pressure gradient, not to the pressure.

\subsection{Reduced POP-$\alpha$ barotropic implementation \label{s_red_alpha}}
The full algorithm, (\ref{full_eta}---\ref{full_U}), is an exact derivation of the LANS-$\alpha$ primitive equations, but has the overwhelming disadvantage that it is extremely slow due to the Helmholtz inversion on the LHS of (\ref{full_eta}).  In this section we test a reduced algorithm that does not include this inversion, and find that numerical experiments of this reduced algorithm are almost identical to the original.

First, we must review the barotropic POP algorithm as it is actually implemented in the code:
\bea
&& \bUh = \bU^{n-1} 
    + \tau 
     \left[ {\bf G}^n- {\sf B}\bU^n
   -\gamma g\del\left( \eta^n + 2\eta^{n-1} \right) 
   \right] \label{POP_Ua} \\
&& \left( \del\cdot H \del
       - \frac{2}{\gamma g\tau^2}\right)
       \eta^{n+1} =  
   - \frac{2}{\gamma g\tau^2}\eta^n 
   +  \del\cdot H \left( \frac{1}{\tau\gamma g} \bUh
    + \del \eta^{n-1} \right)
\\
&& \bU^{n+1} = \bUh - \tau\gamma g 
       \del\left( \eta^{n+1} - \eta^{n-1} \right). \label{POP_U}
\eea
These equations are equivalent to (\ref{cont_stPOP_expC3}---\ref{momx_stPOP_expC1b}); they have the additional step of computing an auxiliary velocity $\bUh$ because it is required when the Coriolis term is implicit.

Our goal in designing the reduced POP-$\alpha$ algorithm was to capture the effects of the LANS-$\alpha$ model, as represented by the full algorithm, but with as few additional computational steps as possible.  The reduced algorithm,
\boxedeqn{Reduced POP-$\alpha$ barotropic algorithm:}{
&& \bVh = \bV^{n-1} 
    + \tau 
     \left[ {\bf G}^n- {\sf B}\bU^n
   -\gamma g\del\left( \eta^n + 2\eta^{n-1} \right) 
   \right]
 \label{red_Va} \\
&& \bUh =  \left(1-\alpha^2\del^2\right)^{-1} \bVh \label{red_Ua} \\
&& \left( \del\cdot H \del
       - \frac{2}{\gamma g\tau^2}\right)
       \eta^{n+1} =  
   - \frac{2}{\gamma g\tau^2}\eta^n 
    + \del\cdot H \left( \frac{1}{\tau\gamma g} \bUh 
    + \del \eta^{n-1} \right)  \label{red_eta}
\\
&& \bV^{n+1} = \bVh - \tau\gamma g 
       \del\left( \eta^{n+1} - \eta^{n-1} \right)  \label{red_V}
\\
&& \bU^{n+1} = \bUh - \tau\gamma g 
       \del\left( \eta^{n+1} - \eta^{n-1} \right),  \label{red_U}
}
uses only a single smoothing step, which is in (\ref{red_Ua}).  In terms of equations, there are only two additions to the original POP implementation (\ref{POP_Ua}---\ref{POP_U}): the computation of the smooth auxiliary velocity in (\ref{red_Ua}) and the final smooth velocity in (\ref{red_U}).


The reduced algorithm (\ref{red_Va}---\ref{red_U}) is written with auxiliary velocities $\bUh,\bVh$ because the POP code is structured in this way to accommodate an implicit Coriolis force.  However, we are using an explicit Coriolis force, so the algorithm can be rewritten in an equivalent form without the auxiliary velocities as:
\bea
&&   \left( \del\cdot H \del
             -\frac{2}{\gamma g\tau^2} \right)\eta^{n+1}
 =   -\frac{2}{\gamma g\tau^2} \eta^n
   + \frac{1}{\tau\gamma g}\nabla\cdot H
    \left(1-\alpha^2\del^2\right)^{-1}\bV^{n-1} 
\nn\\&& \;\;\;\;\;\;
   + \frac{1}{g\gamma} \del \cdot H 
       \left(1-\alpha^2\del^2\right)^{-1}\left[
   {\bf G}^n - {\sf B}\bU^n
   -g\gamma\del\left( \eta^n + 2\eta^{n-1} \right) 
   \right]  
\nn\\&& \;\;\;\;\;\;
   + \frac{1}{g\gamma} \del \cdot H \del \eta^{n-1} \label{red_eta_no_aux}\\
&& \bU^{n+1} = \left(1-\alpha^2\del^2\right)^{-1} \bV^{n-1} 
\nn\\&& \;\;\;\;\;\;
    + \tau \left(1-\alpha^2\del^2\right)^{-1} \left[ 
   {\bf G}^n - {\sf B}\bU^n
   -g\gamma\del\left(\eta^n + 2\eta^{n-1} \right) 
   \right]
\nn\\&& \;\;\;\;\;\;
    + \tau  \left[ 
   -g\gamma\del\left(\eta^{n+1} - \eta^{n-1} \right) 
   \right], 
\\
&& \bV^{n+1} =  \bV^{n-1}
     + \tau \left[ {\bf G}^n -{\sf B}\bU^n
   - \gamma g\del\left(\eta^{n+1} + \eta^n + \eta^{n-1} \right)
    \right]. 
\eea

In the standard POP implementation, the Coriolis term in the barotropic momentum equation may be chosen to be implicit or explicit using $\xi'$ and $\gamma'$ in (\ref{def_U_xi}).  The POP-$\alpha$ algorithms presented here use an explicit Coriolis force.  However, standard POP simulations typically use an implicit Coriolis force, which allows one to take a longer timestep (Table \ref{t_timing}).  Thus there is a motivation to implement POP-$\alpha$ with an implicit Coriolis term as well.  Following Appendix C of \cite{Dukowicz_Smith94jgro}, it can be shown that (\ref{red_Va}) should be replaced by
\bea
&& \left({\bf I} + \tau \gamma {\sf B}\left(1-\alpha^2\del^2\right)^{-1} 
      \right)\left(\bVh-\bV^{n-1}\right) = \nn \\ &&
    + \tau \left[{\bf G}^n
    -\gamma {\sf B}\left(\bU^{n} +2\bU^{n-1}\right)   
   - \gamma g\del\left(\eta^{n+1} + \eta^n + \eta^{n-1} \right)
     \right].
\eea
for an implicit Coriolis force.
Unfortunately, solving this equation would be extremely slow since the inversion of 
$\left({\bf I} + \tau \gamma {\sf B}\left(1-\alpha^2\del^2\right)^{-1}\right)$
would require an iterative routine.  In the standard POP code, this operator is simply 
$\left({\bf I} + \tau \gamma {\sf B}\right)$,
 which is a 2x2 matrix and only takes a few operations to solve for each grid point.  If an iterative routine, like a CG solver, is required to implement the POP-$\alpha$ momentum equation with implicit Coriolis, it would negate any efficiency gains due to the smaller time step.  Thus POP-$\alpha$ was not implemented with an implicit Coriolis term.

Fortunately, POP-$\alpha$ with explicit Coriolis runs stably using longer timesteps than standard POP with explicit Coriolis (Table \ref{t_timing}).  In fact, the maximum timestep with POP-$\alpha$ was found to be comparable to that with standard POP and implicit Coriolis.  This is an important benefit of the LANS-$\alpha$ model, and makes the lack of an implicit Coriolis version a moot point.

\subsection{Unstable variations of the reduced POP-$\alpha$ algorithm}

A slightly altered version of the reduced algorithm was tried in which (\ref{red_eta}) was replaced with
\bea
&& \left( \del\cdot H \del
       - \frac{2}{\gamma g\tau^2}\right)
       \eta^{n+1} =  
   - \frac{2}{\gamma g\tau^2}\eta^n 
\nn \\ &&\;\;\;\;\;\;
    + \del\cdot H \left( \frac{1}{\tau\gamma g} \bUh 
    + \left(1-\alpha^2\del^2\right)^{-1}
       \del \eta^{n-1} \right).   \label{red2_eta}
\eea
The only change from (\ref{red_eta}) is that the 
$\del \eta^{n-1}$ term 
is smoothed here.  This version is closer to the full 
algorithm, as 
can be seen by substituting in the auxiliary velocity,
\bea
&&   \left( \del\cdot H \del
             -\frac{2}{\gamma g\tau^2} \right)\eta^{n+1}
 =   -\frac{2}{\gamma g\tau^2} \eta^n
   + \frac{1}{\tau\gamma g}\nabla\cdot H
    \left(1-\alpha^2\del^2\right)^{-1}\bV^{n-1} 
\nn\\&& \;\;\;\;\;\;
   + \frac{1}{g\gamma} \del \cdot H 
       \left(1-\alpha^2\del^2\right)^{-1}\left[
   {\bf G}^n - {\sf B}\bU^n
   -g\gamma\del\left( \eta^n + \eta^{n-1} \right) 
   \right]  \label{red2_eta_no_aux}
\eea
That is, (\ref{red2_eta_no_aux}) is more similar to the full algorithm
(\ref{full_eta}) (only one smoothing, on the LHS, is missing) than is
the reduced algorithm (\ref{red_eta_no_aux}).  Despite being closer in
appearance to the full algorithm, this version of the reduced
algorithm was found to be unstable in practice, and is therefore not a
viable option.  This instability is not revealed by the linear
stability analysis; the damping factor for this algorithm is nearly
identical to the full POP-$\alpha$ algorithm.  One possible
explanation for this instability is that when the smoothing operator
is removed from the LHS of the $\eta$ equation in the reduced
algorithm, it must be partially removed from the RHS as well.

Another alteration of the reduced POP-$\alpha$ algorithm is to replace (\ref{red_U}) with
\bea
&\bU^{n+1} = \left(1-\alpha^2\del^2\right)^{-1}\bV^{n+1}.
\label{red_U_Helm}
\eea
Simulations where this method was employed were often unstable.  Even
though (\ref{red_U_Helm}) follows directly from the LANS-$\alpha$
equations, it appears that (\ref{red_U})--which lacks the smoothing
operation on the $\del \eta$ terms--works better in practice.  Again,
this is probably because the reduced barotropic algorithm lacks
pressure gradient smoothing in (\ref{red_eta}), and a corresponding
lack of smoothing in (\ref{red_U_Helm}) makes for a more stable
algorithm.

One of the difficulties encountered in devising the appropriate
POP-$\alpha$ algorithm was to ensure that the velocity is nondivergent
at each baroclinic level.  The LANS-$\alpha$ equations specify that
the {\it smooth} velocity $\bu$ satisfies the continuity equation
(\ref{Helm1}).  This is a practical requirement as well; simulations
where the rough velocity $v$ is used in the free surface height
equation (\ref{red_eta_no_aux}) (which is derived from the continuity
equation) were found to be unstable.

\section{Stability Analysis of the POP-$\alpha$ Barotropic Solver}

The stability analysis of the POP algorithm by Dukowicz and Smith \cite{Dukowicz_Smith94jgro} was used to find the best weighting for the implicit/explicit variables in (\ref{def_U_xi}--\ref{def_U_theta}), which are $\gamma=\xi=\xi'=1/3$ and $\theta=1$.  For gravity waves, this choice of parameters strongly damps the computational mode, and slightly damps and slows down the the physical modes.  For Rossby waves, the computational mode is slightly damped, and the physical mode is nearly undamped but slowed down.  These results are reproduced here, in order to compare with the stability analysis of POP-$\alpha$.  The analysis is partly based on previous work by Wingate \cite{Wingate04mwr}, who investigated the stability of the LANS-$\alpha$ shallow water equations by comparing a third-order Adams-Bashforth method to the Dukowicz and Smith \cite{Dukowicz_Smith94jgro} POP barotropic algorithm.

The barotropic component of POP (and POP-$\alpha$) compute the vertically integrated velocities and the free surface height, and is therefore the same as the shallow water equations.  The dispersion relation for the continuous equations is derived by transforming the equations of motion into dimensionless form, combining them, and assuming a plane wave solution of the form $e^{i(k_h\cdot{\bf x}-\omega t)}$ for the remaining variable.  This analysis appears in \cite{Dukowicz_Smith94jgro} and \cite{Wingate04mwr}, and only the results are stated here.
For the continuous shallow water equations, the dispersion relation for (external) gravity waves are 
\bea
\omega_g = \frac{k_h}{F^2}
\eea
for the shallow water equations (used by the barotropic component of standard POP) and
\bea
\omega_g = \frac{k_h}{F^2\left(1+\alpha^2 k_h^2\right)}
\eea
for LANS-$\alpha$ (used in POP-$\alpha$), where $k_h^2=k^2+l^2$ is the horizontal wave number,  $F=U/\sqrt{gH_o}$ is the Froude number, $g$ is gravitational acceleration, $H_o$ is the average fluid depth, and $U$ a typical velocity scale.  
The dispersion relation for Rossby waves is 
\bea
\omega_r = \frac{-k\beta'}{k_h^2 + 1/B^2}  
\label{dr_RW_NS}
\eea
for the shallow water equations and
\bea
\omega_r = \frac{-k\beta'}{k_h^2\left( 1+ \alpha^2 k_h^2\right) + 1/B^2}
\label{dr_RW_alpha}
\eea
for LANS-$\alpha$, 
where 
$B=R/L$, $R=\sqrt{gH_o}/f_o$ is the Rossby deformation radius,  $f=f_0 + \beta y$ is the Coriolis parameter, 
$\beta'=\beta L^2/U$ is the dimensionless beta parameter, $\beta=\partial_y f$, and $L$ is a typical length scale.
For scales that are much larger than alpha ($1/k_h>>\alpha$), $1+\alpha^2 k_h^2\rightarrow 1$, and the LANS-$\alpha$ dispersion relations are identical to their Navier-Stokes counterparts.  For scales near alpha, $1+\alpha^2 k_h^2\sim 2$, so that both the gravity waves and Rossby waves are slowed down.

We now investigate how varying $\alpha$ effectively changes the Rossby radius in these equations.  First, define the Rossby wavenumber as $k_r=L/R$, and note that the Rossby wavenumber maximizes the dispersion relation for the Navier-Stokes equation (\ref{dr_RW_NS}).  This is clear in plots of the dispersion relation (see Fig. 2 in \cite{Wingate04mwr}).  Analogously, we define an effective Rossby radius $R^*$ for the LANS-$\alpha$ model using the wavenumber that maximizes (\ref{dr_RW_alpha}).  Solving $d\omega_r/dk=0$ for $k$ and letting $l=0$ and $k=L/R^*$, we produce a relationship between the Rossby radius $R$ and the effective Rossby radius $R^*$ as a function of alpha:
\bea
3\left( \frac{\alpha}{R} \right)^2 
\left( \frac{R}{R^*} \right)^4
+\left( \frac{R}{R^*} \right)^2
-1 = 0.
\eea
When $\alpha=0$ (no alpha model) then $R=R^*$, as expected.  As $\alpha$ increases, the effective Rossby radius increases (Fig. \ref{f_Rossby_radius}).  This is the mechanism by which POP-$\alpha$ allows more eddy activity than standard POP at scales just above the actual Rossby radius.

\subsection{Gravity waves}
In this section we conduct the stability analysis for gravity waves in the discrete barotropic equations for standard POP, full POP-$\alpha$, and reduced POP-$\alpha$.  The notation of the derivation follows \cite{Wingate04mwr}, section 4.2.  The starting point is the POP-$\alpha$ barotropic equations, (\ref{mom_POPa}--\ref{Helm_btrop}), which are then rescaled using a velocity scale $U$, a length scale $L$, a height scale $H_0$, and a time scale $\tau$.  In the limit where the timescale $\tau=L/U$, the equations are 
\bea
\delta^{n+1}-\delta^{n-1} + \frac{2\Delta t}{F^2} \del^2 \eta^{\xi} = 0, 
   \label{gw_div}\\
\eta^{n+1}-\eta^n + \Delta t \deltab^{\theta} = 0,\\
\deltab^{n+1}= \left(1-\alpha^2 \del^2 \right)^{-1} \delta^{n+1} 
\label{Helm_delta}
\eea
where $\delta = \p_x U_1 + \p_y U_2$ is the divergence, and the superscripts $\xi$ and $\theta$ have the same meaning as (\ref{def_eta_xi}-\ref{def_U_theta}).  The standard POP equations are recovered when $\alpha\rightarrow 0$ so that $\deltab=\delta$.   Substituting 
$\delta^n=\lambda^n e^{ik_hx}{\hat \delta}$ and
$\eta^n=\lambda^n e^{ik_hx}{\hat \eta}$ into the above equations, we obtain the characteristic polynomial
\bea
\left( \lambda^2 - 1 \right)
\left( \lambda   - 1 \right)
+ \frac{2Q^2}{F^2} P S
= 0. \label{cp_gw}
\eea
\cite{Dukowicz_Smith94jgro,Wingate04mwr} where $Q^2=\Delta t^2 k_h^2$, $Q$ is a nondimensional CFL number, and 
\bea
S= \theta \lambda + 1 - \theta. \label{S}
\eea

The difference between POP, full POP-$\alpha$, and reduced POP-$\alpha$ in this analysis is in the polynomial $P$, which comes from the $\del^2 \eta^{\xi}$ term.  For POP,
\bea
P=\xi\lambda^2 + (1-\xi-\gamma)\lambda + \gamma. \label{P}
\eea
For the {\it full} POP-$\alpha$ algorithm, a Helmholtz inversion is applied to the full  $\del \eta^{\xi}$ term in (\ref{full_eta}), so that
\bea
P=\left(1+\alpha^2 |k_h|^2 \right)^{-1}
  \left(\xi\lambda^2 + (1-\xi-\gamma)\lambda + \gamma\right).
\eea
For the {\it reduced} POP-$\alpha$ algorithm, the Helmholtz inversion is not applied to $\del \eta^{n+1}$ and one $\del \eta^{n-1}$ in (\ref{red_eta}):
\bea
P=\left(\xi\lambda^2 + \left(1+\alpha^2 |k_h|^2 \right)^{-1}
\left((1-\xi-\gamma)\lambda + 2\gamma\right) -\gamma\right). \label{P2}
\eea

\begin{figure}[tbh]
(a)\scalebox{.9}{\includegraphics{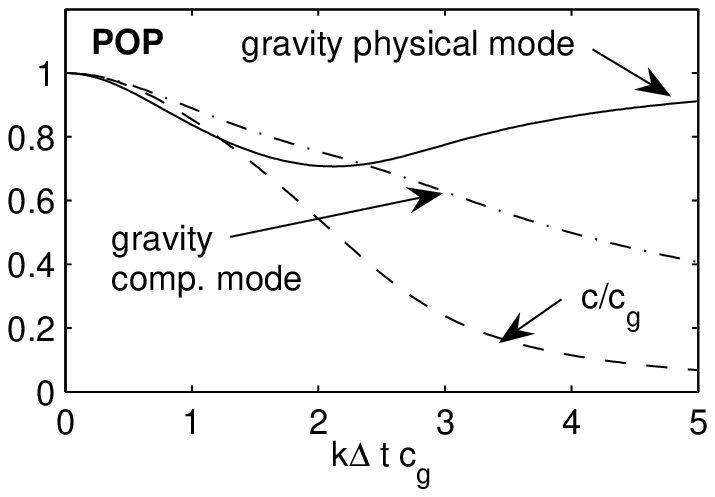}} 
(b)\scalebox{.9}{\includegraphics{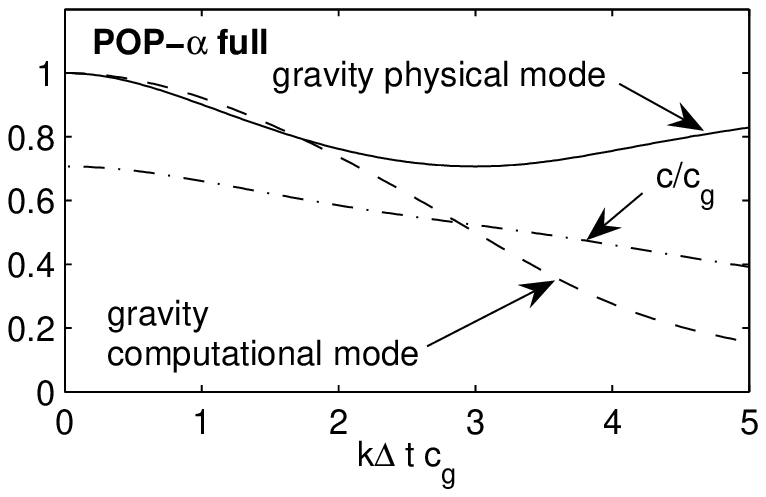}} \\
(c)\scalebox{.9}{\includegraphics{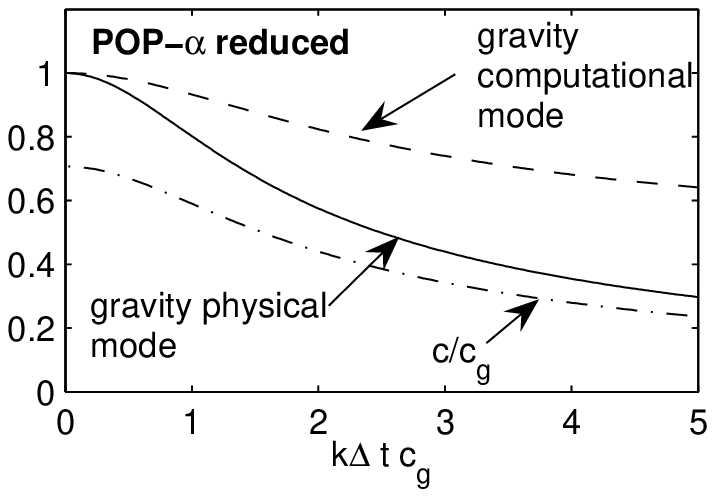}} 
\caption{\label{f_gw_st_an} 
Gravity wave stability analysis results for POP (a), POP-$\alpha$ full algorithm (b), and POP-$\alpha$ reduced algorithm (c).
Each plot shows the damping factors and phase speed as a function of the Courant-Friedrichs-Lewy number $k\Delta t c_g$, and uses $\alpha=1/k_h$, $\xi=\gamma=1/3$, and $\theta=1$.
}\end{figure}

\begin{figure}[tbh]
\scalebox{.9}{\includegraphics{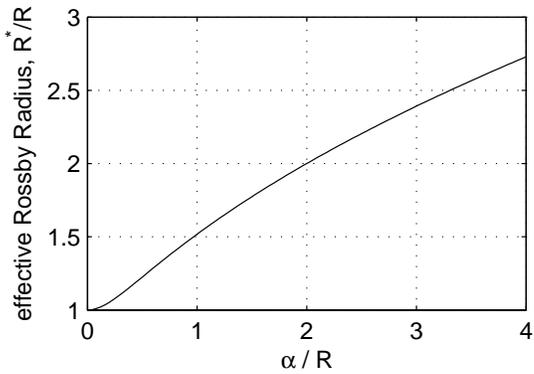}} 
\caption{\label{f_Rossby_radius} 
The effective Rossby radius of deformation, $R^*$, as a function of $\alpha$, where both are normalized by the non-alpha model Rossby Radius, $R$.  This shows that the LANS-$\alpha$ model increases the effective Rossby Radius.  This relation was produced using the dispersion relation for Rossby waves in the continuous shallow water LANS-$\alpha$ equations.  The effective Rossby deformation radius
includes the effects of the small, unresolved  scales on the large.
}\end{figure}

The gravity wave amplification factors and dispersion errors for these algorithms are shown in Fig. \ref{f_gw_st_an}.  These curves were computed numerically from the characteristic polynomial (\ref{cp_gw}) using typical parameters, $\theta=1$ and $\xi=\gamma=1/3$.  The Coriolis term does not appear in the divergence equation (\ref{gw_div}), so this analysis is valid for both implicit and explicit Coriolis schemes, and the parameters $\xi'$ and $\gamma'$ do not appear.  Because the plots are shown with $Q$ on the horizontal axis, $\alpha|k_h|$ is left as a free parameter and was chosen such that $\alpha\sim 1/|k_h|$.  

Fig. \ref{f_gw_st_an} shows that all schemes are stable to linear gravity waves, because all of the damping factors are less than one.  Fig. \ref{f_gw_st_an}a, for standard POP, is the basis of comparison, and is identical to Fig. 2 in \cite{Dukowicz_Smith94jgro}.  The POP-$\alpha$ reduced algorithm damps the physical gravity waves more strongly and the computation gravity waves less strongly than both POP and the full POP-$\alpha$ algorithm.


\subsection{Rossby waves}
To conduct a stability analysis of the Rossby waves, we begin with the POP-$\alpha$ equations in streamfunction-divergence form,
\bea
Ro\left(\del^2 \Psi^{n+1} - \del^2 \Psi^{n-1} \right)
+ 2 \Delta t \left( \beta'Ro \Psib_x^{\xi'} + \deltab^{\xi'} \right) = 0, 
\label{st_an_rw_vort}\\
Ro\left(\Psib^{n+1} - \Psib^{n}\right) + \Delta t B^2 \deltab^\theta=0,\\
U_1^n=-\p_y\Psib^n,\;\;\;
U_2^n= \p_x\Psib^n,\;\;\;
\eta^n=\Psib^n,\\
\Psib^{n}= \left(1-\alpha^2 \del^2 \right)^{-1} \Psi^{n}.
\eea
(\cite{Wingate04mwr}, sections 1b2 and 5b), where the Rossby number $Ro=U/f_oL$.  Again, the standard POP equations are recovered when $\alpha\rightarrow 0$, so that $\Psib=\Psi$ and $\deltab=\delta$.  The free surface height $\eta$ is equal to the {\it smooth} streamfunction $\Psib$ because of geostrophic balance, where the $\del \eta$ is balanced by the Coriolis force, which uses the smooth velocity.  Geostrophic balance also requires that the implicit/explicit weighting of the Coriolis velocity (\ref{def_U_xi}) and pressure (\ref{def_eta_xi}) are identical, so that $\xi=\xi'$ and $\gamma=\gamma'$ (\cite{Dukowicz_Smith94jgro}, section 3.3).  This means that the explicit Coriolis formulation, where $\xi'=\gamma'=0$, cannot be considered in this stability analysis, because for stability $\xi>1/4$ is required \cite{Dukowicz_Smith94jgro}.

This system has a characteristic polynomial of
\bea
\left(1+\alpha^2 |k_h|^2 \right)
B^2k_h^2\left( \lambda^2-1 \right)
S 
 + 2P
\left[ \lambda - 1 - \Delta tik\beta' B^2
S \right] 
=0 \label{cp_rw}
\eea
where $S$ is defined in (\ref{S}) and 
\bea
P=\xi\lambda^2 + (1-\xi-\gamma)\lambda + \gamma.
\eea

In this Rossby wave analysis, the full and reduced POP-$\alpha$ algorithms have identical characteristic polynomials.  That is because when one takes the curl of the momentum equation to get the vorticity equation (\ref{st_an_rw_vort}), the $\del \eta$ term drops out, and the $\del \eta$ term is where the differences between the full and reduced POP-$\alpha$ algorithms appear.  

\begin{figure}[tbh]
(a)\scalebox{.9}{\includegraphics{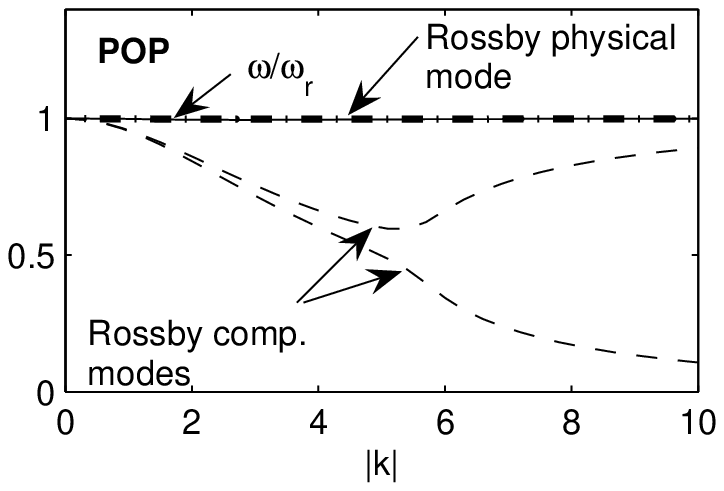}}
(b)\scalebox{.9}{\includegraphics{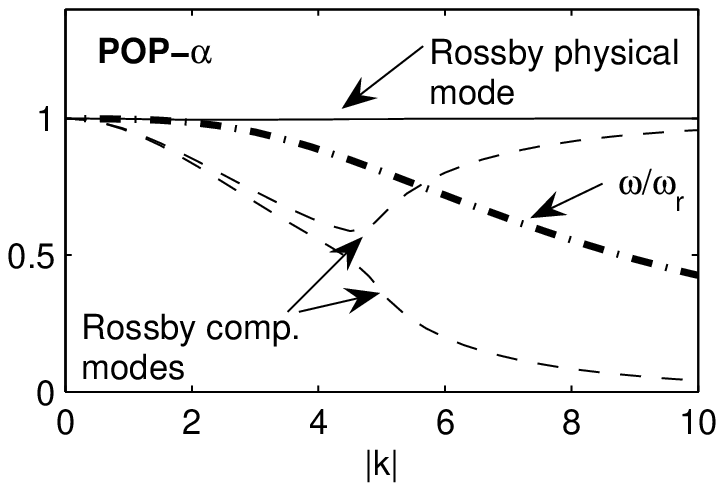}} 
\caption{\label{f_rw_st_an} 
Rossby wave stability analysis results for POP (a) and POP-$\alpha$ (b) as a function of wave number $k$.  For both algorithms, the physical Rossby wave is undamped.  POP-$\alpha$ slows down Rossby waves, while POP does not.  Here $B^2=1/4$ and $\alpha=1/8$ of the domain width.
}\end{figure}

Here $B^2=R^2/L^2$, so the alpha model is effectively making the Rossby Radius, $R$, larger in the first term of (\ref{cp_rw}).  This same effect was observed in the continuous equations (Fig. \ref{f_Rossby_radius}).  The damping factors, using the typical choice of parameters ($\theta=1, \xi=\gamma=\xi'=\gamma'=1/3$) are shown in Figure \ref{f_rw_st_an}.  Compared to standard POP, POP-$\alpha$ slows down the Rossby waves.  The damping of Rossby waves is unaffected in this analysis.


\subsection{Full Stability Analysis}

A stability analysis of both gravity and Rossby waves may be conducted by beginning with the full beta-plane equations in vorticity-divergence, dimensionless form (following \cite{Dukowicz_Smith94jgro}, equations 19 and 41)
\bea
\p_t \delta - {\bar \zeta} + \epsilon U_1 + \del^2\eta=0,\\
\p_t\zeta + \deltab + \epsilon U_2=0,\\
\p_t\eta + \deltab=0
\eea
where the barred variables are smoothed, as in (\ref{Helm_delta}), and $\epsilon=\beta L/f_o$.  Discretizing and introducing Fourier modes as before, one obtains a fifth-degree complex characteristic polynomial for the amplification factors:
\bea
&& 2|k_h|^4\Delta t^2(\lambda^2-1) P S
-4i\epsilon k |k_h|^2 \Delta t^3 P P' S \nn\\
&& + |k_h|^2 (\lambda-1)(\lambda^2-1)^2
+4 |k_h|^2 \Delta t^2(\lambda-1)P'^2 \nn\\
&& -2i\epsilon k \Delta t (\lambda-1)(\lambda^2-1) P'
-4i\epsilon l \Delta t^2(\lambda-1)P'^2=0
\eea
where $S$ is defined in (\ref{S}), 
\bea
P'=\xi'\lambda^2 + (1-\xi'-\gamma')\lambda + \gamma'
\eea
depends on the implicit/explicit Coriolis parameters, and 
$P$ is defined as in (\ref{P}--\ref{P2}) for POP and various POP-$\alpha$ algorithms.

\begin{figure}[tbh]
(a)\scalebox{.9}{\includegraphics{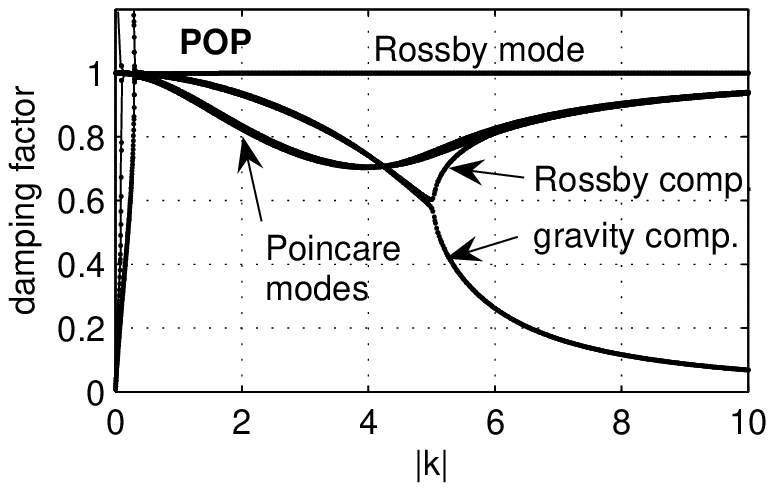}} 
(b)\scalebox{.9}{\includegraphics{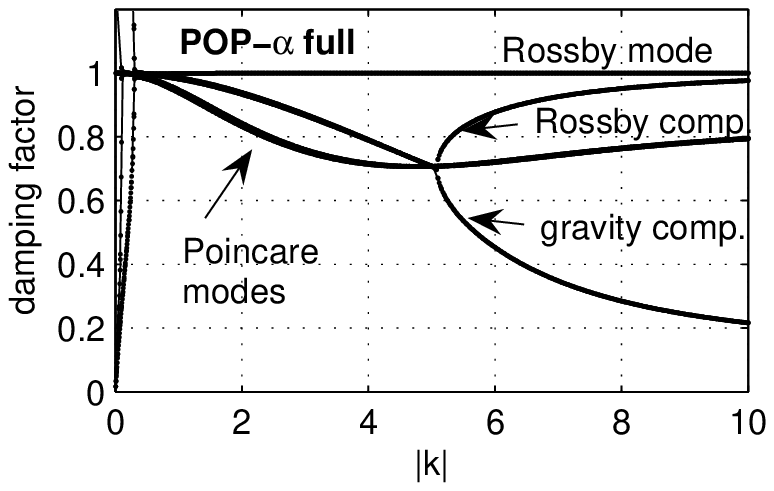}} \\
(c)\scalebox{.9}{\includegraphics{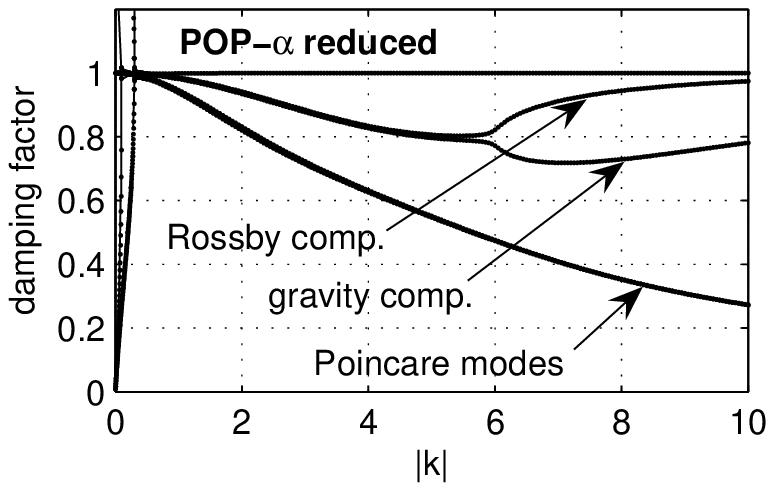}} 
\caption{\label{f_st_an} 
Damping factor for the discrete equations for purely zonal waves of the barotropic beta plane equations at midlatitudes ($\epsilon=3$), using a time step of $\Delta t f_o=1/2$, $\alpha=1/8$ of the domain width, and typical POP parameters ($\xi=\gamma=1/3$, $\theta=1$) and explicit Coriolis terms ($\xi'=\gamma'=0$).
}\end{figure}

Figure \ref{f_st_an} shows that all versions of POP are stable when $\Delta t f_o=1/2$ and $|k|\ge 1$.  These plots show instabilities when $|k|<1$, but this does not affect the overall stability.  As explained in \cite{Dukowicz_Smith94jgro} for standard POP with explicit Coriolis terms, there is always a region where $|\lambda|>1$ for sufficiently small $|k|$.  In practice, this instability is not an issue if $\Delta t$ is chosen such that $\Delta t f_o < 1$.  This is true of the POP-$\alpha$ algorithms as well.  

Comparing the POP-$\alpha$ reduced algorithm with POP in Fig. \ref{f_st_an}, we see that the Rossby waves are not damped in either, the computational modes are less damped and the Poincare-wave modes are more damped in the POP-$\alpha$ reduced algorithm than in standard POP.

\section{Results \label{s_results}}

Long-time simulations of POP, full POP-$\alpha$, and reduced
POP-$\alpha$ were run in a channel-model domain. In this section, we
show that the reduced POP-$\alpha$ algorithm produces results that are
nearly identical to the full POP-$\alpha$ algorithm, but is much
faster.  For the POP-$\alpha$ simulations, smoothing was achieved by
both a Helmholtz inversion, as prescribed by the LANS-$\alpha$
equations, and a simple filter that averages nearest neighbors.  Using
the filter to smooth also produces a substantial speed-up over the
Helmholtz filter, which requires an iterative method; a comparison of
smoothing methods is addressed in \cite{Petersen07jcp}.

\subsection{Description of the Model Problem}

The model problem can be thought of as an idealization of the
Antarctic Circumpolar Current. The Circumpolar region is unique in the
World Ocean in being zonally continuous, the only region of the ocean
where there is no continent against which a zonal pressure gradient
can be supported, and consequently the only region where meridional
heat transport falls so heavily to the mesoscale eddies. A reentrant
channel model therefore provides a relevant setting in which to
consider the impact of a turbulence parameterization in an ocean
general circulation model. Our test problem is based in part on the
works of Karsten et al. \cite{Karsten_ea02jpo} and Henning and Vallis
\cite{Henning_Vallis05jpo}; the physical analysis in both of those
works is more thorough than what we present, as our focus is on model
development rather than physical oceanography.

\begin{figure}[tbh]
\center
\scalebox{1}{\includegraphics{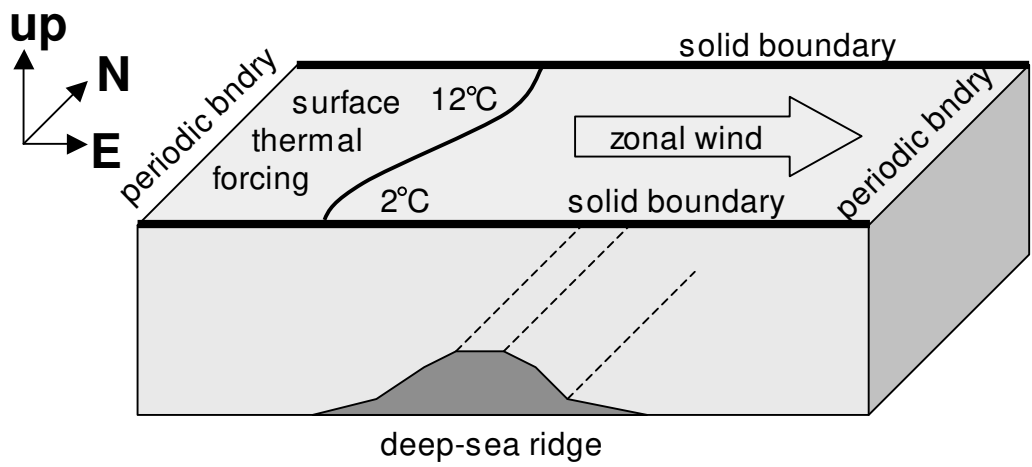}} 
\caption{\label{f_channel_image} 
Schematic of the model domain, which has periodic zonal boundaries, solid north/south boundaries, a deep-sea ridge, surface wind forcing,  and thermal forcing.
}\end{figure}

\begin{figure}[tbh]
\begin{tabular}[c]{cc}
\scalebox{.5}{\includegraphics{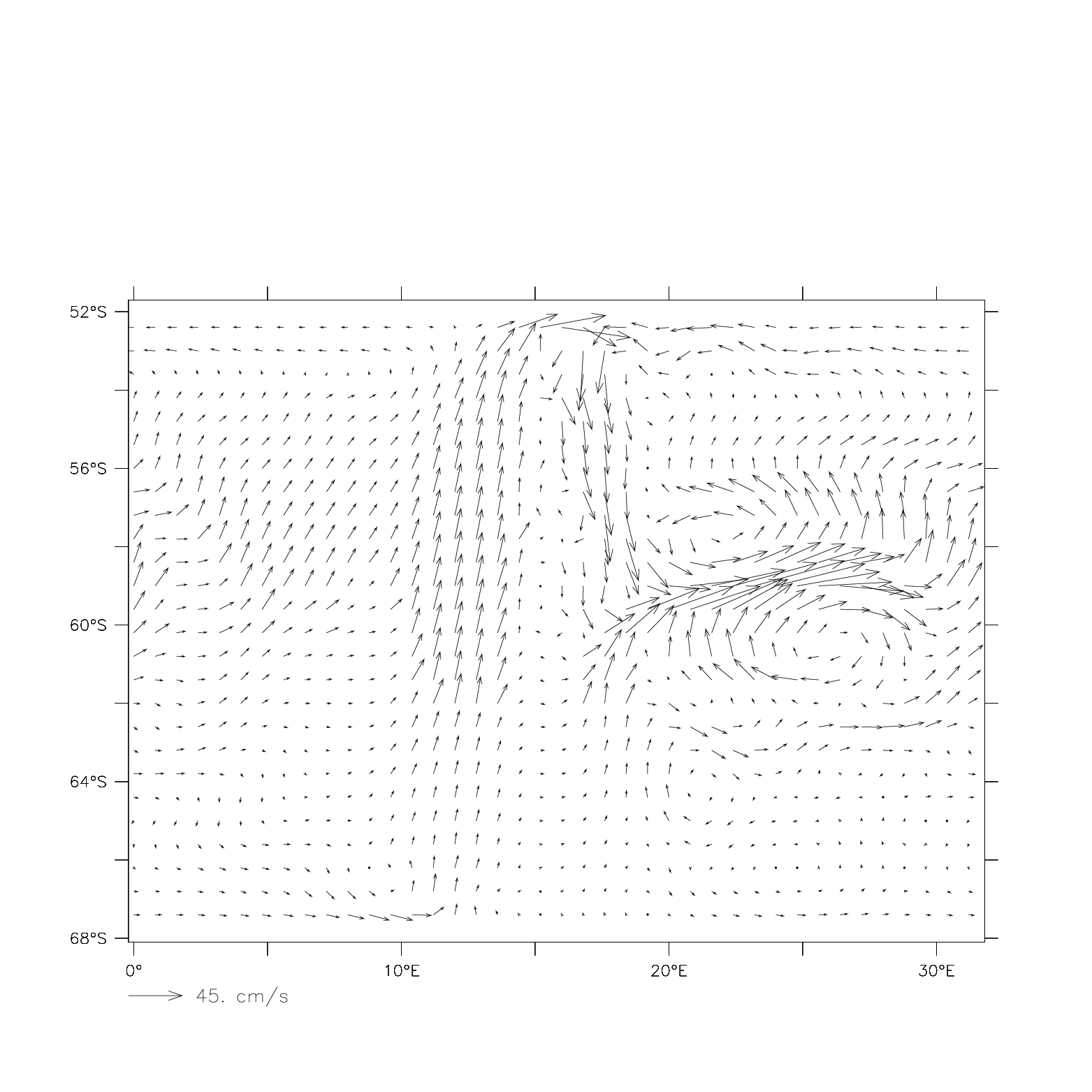}} &
\scalebox{.5}{\includegraphics{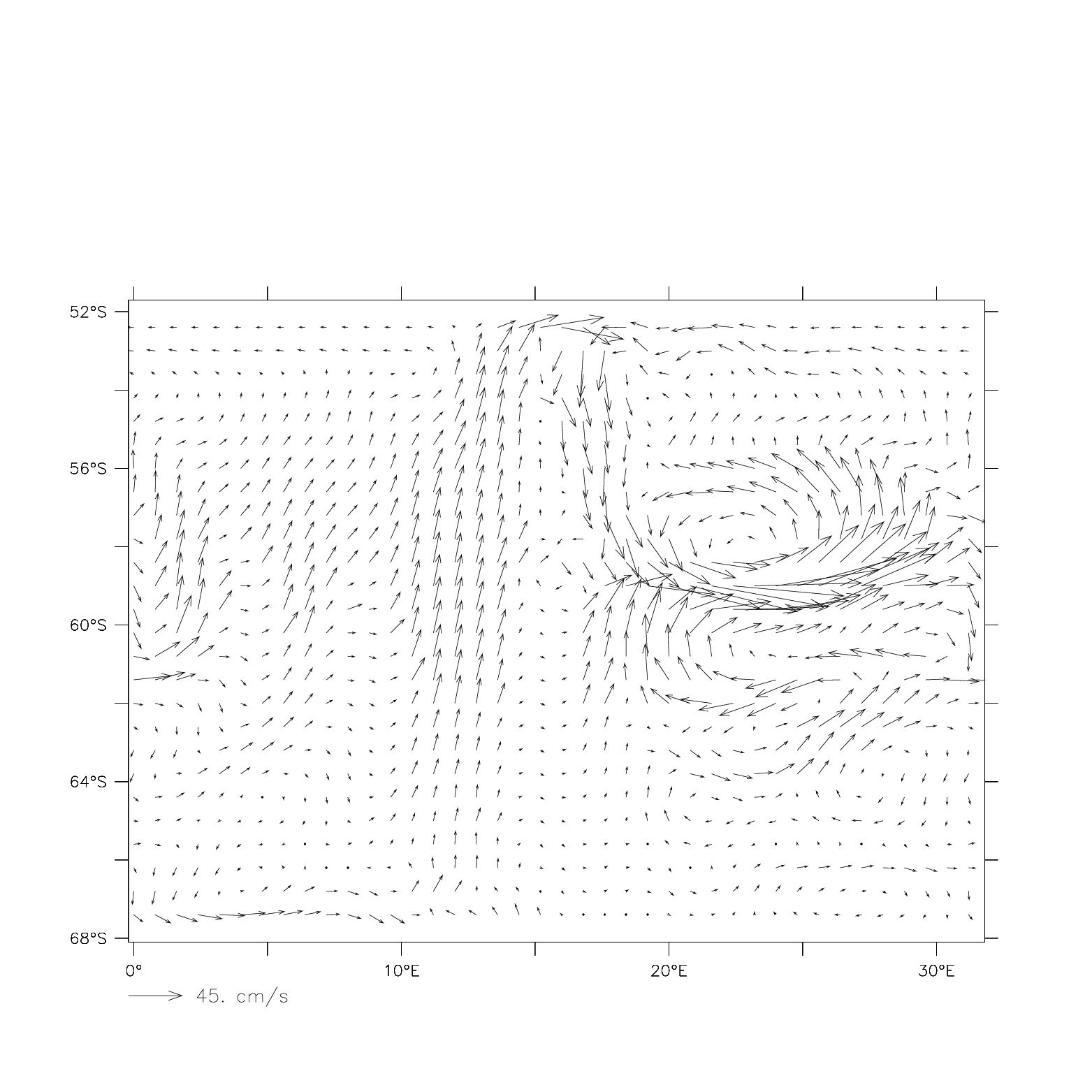}}\\
(a) Surface pot. temp., full algorithm &
(b) Surface pot. temp., reduced algorithm \\
\scalebox{.5}{\includegraphics{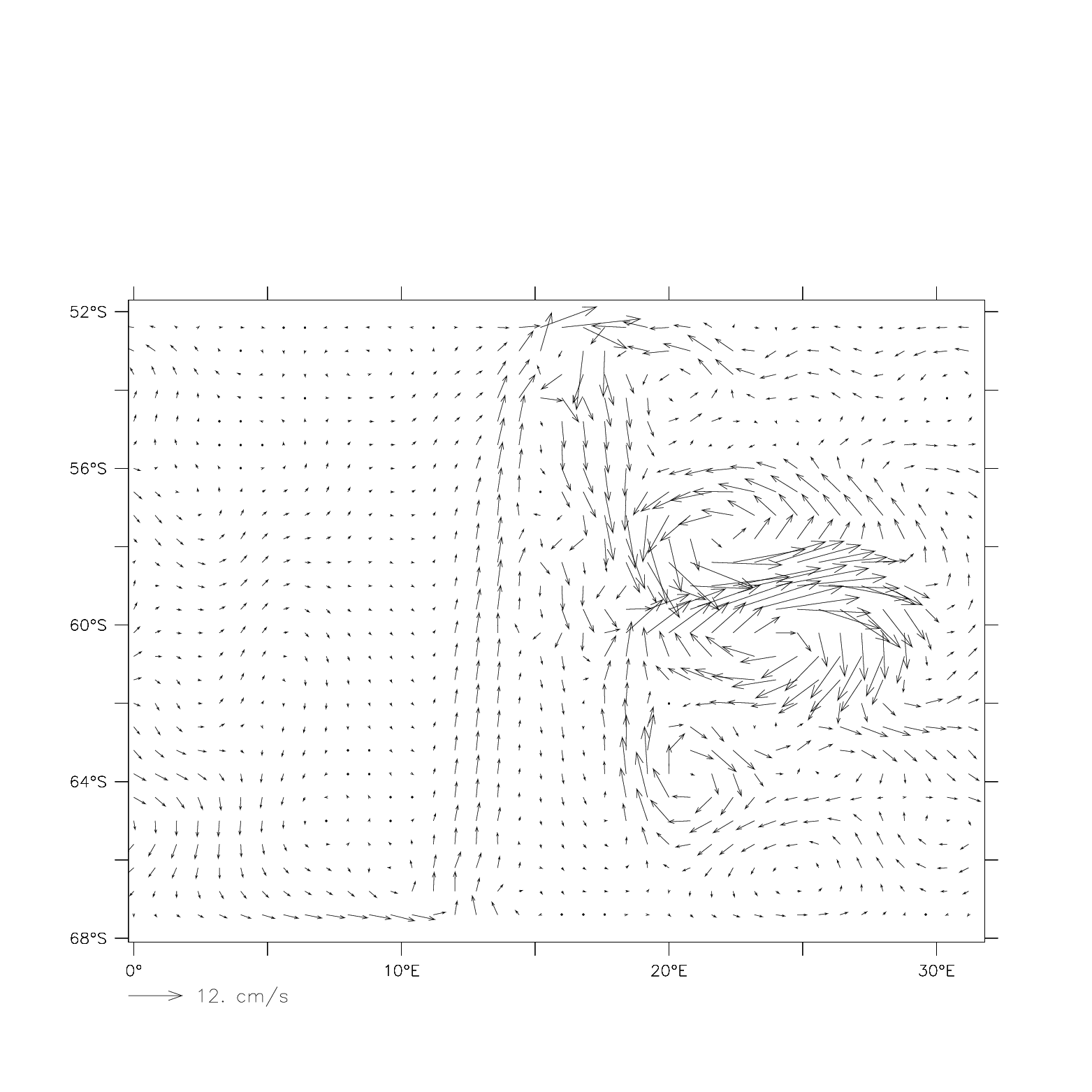}} &
\scalebox{.5}{\includegraphics{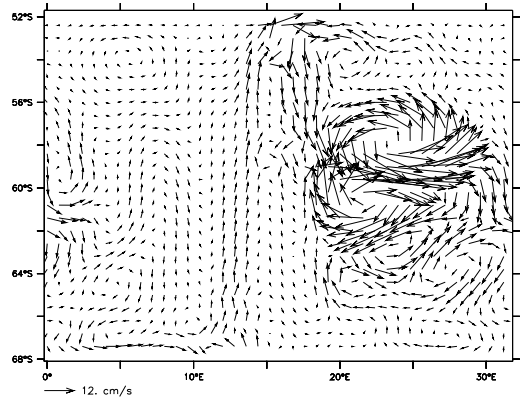}}\\
(c) Pot. temp. at 1600m, full algorithm &
(d) Pot. temp. at 1600m, reduced algorithm 
\end{tabular}
\caption{\label{f_vel_sections} 
Snapshots of the rough velocity field $\bv$ at 150 years for experiment 0.4F.  A deep-sea ridge between 11$^o$E and 18$^o$E causes northward and then southward flow, and spurs eddies east of 18$^o$E.  The full and reduced POP-$\alpha$ algorithm produce dynamics that are nearly identical.
}\end{figure}

The zonally periodic model domain has solid boundaries to the north
and south (Fig. \ref{f_channel_image}).  An eastward wind stress
drives an eastward circulation in the channel.  A deep-sea ridge
between 11E and 18E that is uniform from north to south forces the
water column northward from 10-15$^o$E and then southward again from
15-20$^o$E by conservation of potential vorticity
(Fig. \ref{f_vel_sections}) (see, e.g. \cite{Holton92bk} p. 100).
This deflection of the mean flow spawns mesoscale eddies to the east
of 18$^o$E if the resolution is sufficiently high.  For this study,
POP was run in three resolutions, referred to as 0.8, 0.4, and 0.2 to
correspond with the longitudinal resolution, as shown in Table
\ref{t_parameters}.  The longitudinal resolution was chosen to have an
aspect ratio of one at the central latitude of $60^\circ$ south. At
the lowest resolution, 0.8, the Rossby Radius of deformation is not
resolved, and so the velocity field does not contain eddies.  At the
next higher resolution of 0.4 eddies form, and even finer and more
numerous eddies can be seen in simulation 0.2.

\begin{figure}[tbh]
\center
\scalebox{.8}{\includegraphics{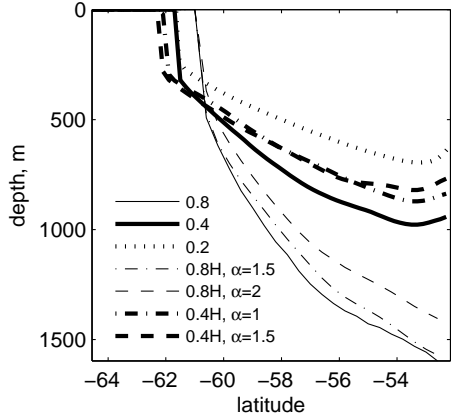}} 
\caption{\label{f_isotherms} 
Depth of 6$^o$C isotherm of potential temperature, averaged between 0$^o$E and 10$^o$E.  Isotherms flatten with increasing resolution of standard POP (solid, dotted), due to the effects of mesoscale eddies.  Simulations using POP-$\alpha$ (dash, dash-dot) have flatter isotherms than standard POP at the same resolution.
}\end{figure}

The model induces a surface thermal forcing by restoring the SST to a
smooth temperature profile ranging from 2$^o$C at 68$^o$S to 12$^o$C
at 52$^o$S.  The thermal forcing in conjunction with the wind stress
drive downwelling of warmer waters in the north and deep penetration
of colder waters in the south, giving rise to the sloping isotherms
seen in Figure \ref{f_isotherms}.  These tilted isotherms are a source
of potential energy, driving baroclinic instability. The mesoscale
eddies generated from this conversion of potential to kinetic energy
tend to flatten the isotherms.

\begin{figure}[tbh]
\center
(a)\scalebox{.7}{\includegraphics{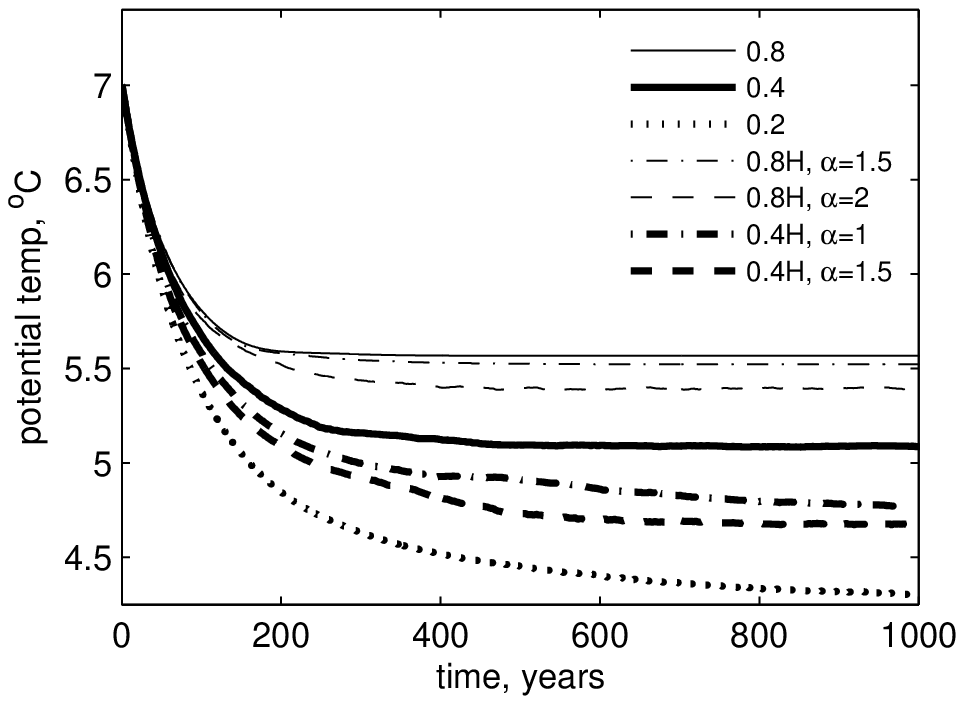}} 
(b)\scalebox{.7}{\includegraphics{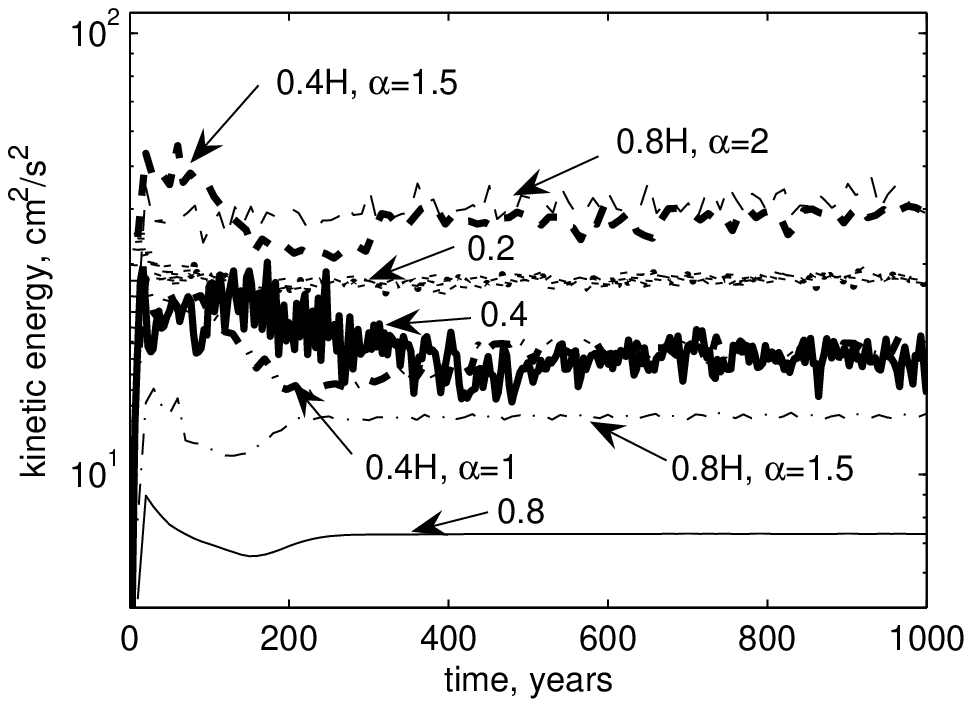}}
\caption{\label{f_T_KE} 
Global mean potential temperature (a) and kinetic energy (b) as a function of time using standard POP and POP-$\alpha$.  Higher resolution simulations (0.4 and 0.2) reach a cooler steady-state (a) and have higher kinetic energy (b) due to the activity of mesoscale eddies.  POP-$\alpha$ simulations (0.8H, 0.4H) caputures the effects of these eddies at lower resolution that standard POP.  This trend increases with larger $\alpha$.
}\end{figure}

The action of the eddies can be gauged through their effect on the
temperature distribution.  As the resolution of standard POP
simulation is increased, mesoscale eddies are better resolved, and the
isotherms are flatter.  Figure \ref{f_isotherms} shows an important
property of the LANS-$\alpha$ model: it allows more eddy activity.  By
capturing the effects of these eddies, lower resolution POP-$\alpha$
simulations also have flatter isotherms than standard POP at the same
resolution.  A global statistic that represents the tilting of
isotherms is potential temperature averaged over the entire domain
(Fig. \ref{f_T_KE}a).  With progressively higher resolution
simulations of standard POP (0.8, 0.4, 0.2) the ocean cools faster and
levels out to a cooler equilibrium.  All simulations begin with a
constant temperature of 7$^o$C.  The cooler global temperature of
higher resolution simulations indicate that the isotherms are more
level due to the eddies.  Higher-resolution effects are also seen in
the kinetic energy using POP-$\alpha$: kinetic energy averaged over
the domain increases with resolution (Fig. \ref{f_T_KE}b); lower
resolution POP-$\alpha$ simulations also capture this effect.

Thus the POP-$\alpha$ results in this channel test problem are similar to those in other numerical simulations of LANS-$\alpha$: they produce turbulence statistics that resemble those from higher-resolution simulations without LANS-$\alpha$\cite{Chen_ea98prl,Chen_ea99pd_b,Holm_Nadiga03jpo}.  Some of these effects, like the flattening of isotherms, must be due to the inclusion of LANS-$\alpha$ in the baroclinic component of POP.  Based on previous work in barotropic LANS-$\alpha$ simulations \cite{Holm_Nadiga03jpo}, barotropic effects are most likely involved as well.

\subsection{Comparison of algorithms \label{ss_results_compare}}

\begin{figure}[tbh]
(a)\scalebox{.7}{\includegraphics{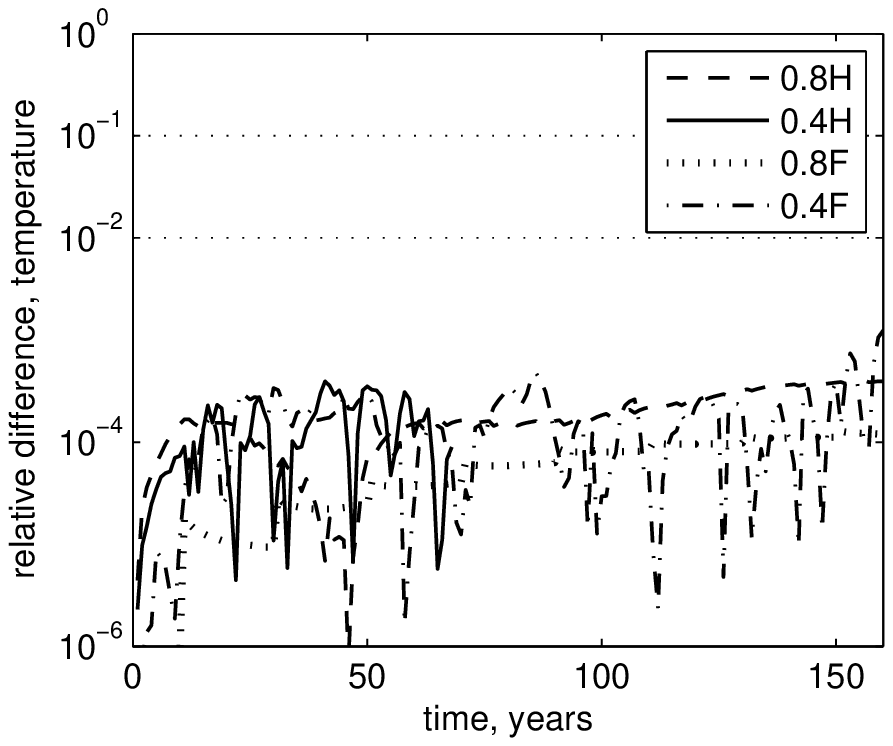}} 
(b)\scalebox{.7}{\includegraphics{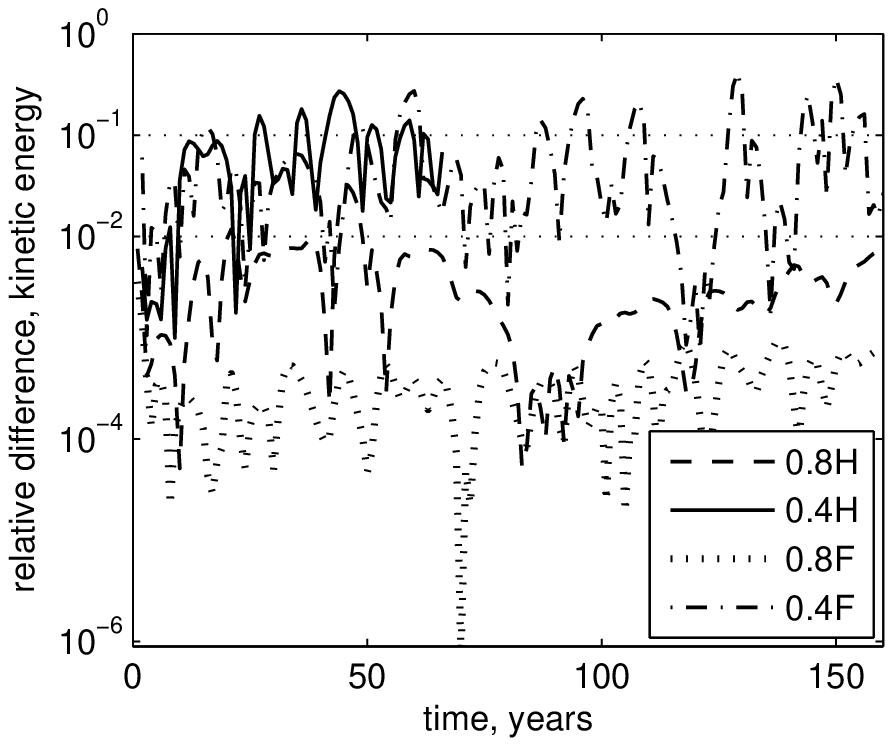}}
\caption{\label{f_T_KE_diff} 
Relative difference between the full POP-$\alpha$ algorithm and reduced POP-$\alpha$ algorithm for several experiments.  The difference in the global mean potential temperature (a) is less than 0.05\% in all cases; the difference in global mean kinetic energy (b) is generally less than 1\% for lower resolution ($0.8^o$) cases, but is larger for higher resolution due to the variability, as seen in Fig. \ref{f_T_KE}b.  Statistics are only available for the first 50 years of 0.4H because the full POP-$\alpha$ algorithm is so slow. 
}\end{figure}

\begin{figure}[tbh]
\begin{tabular}[c]{cc}
\scalebox{.5}{\includegraphics{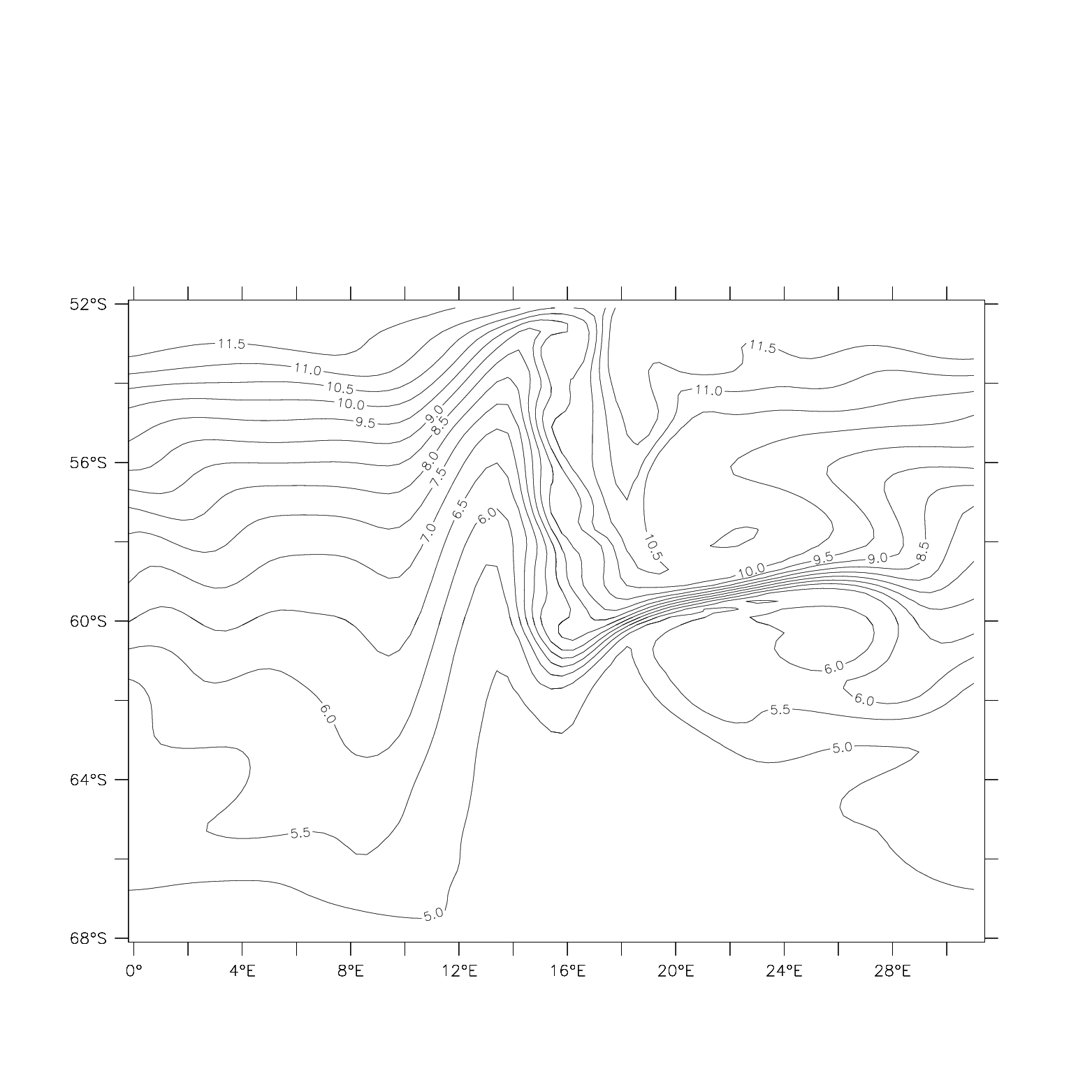}} &
\scalebox{.5}{\includegraphics{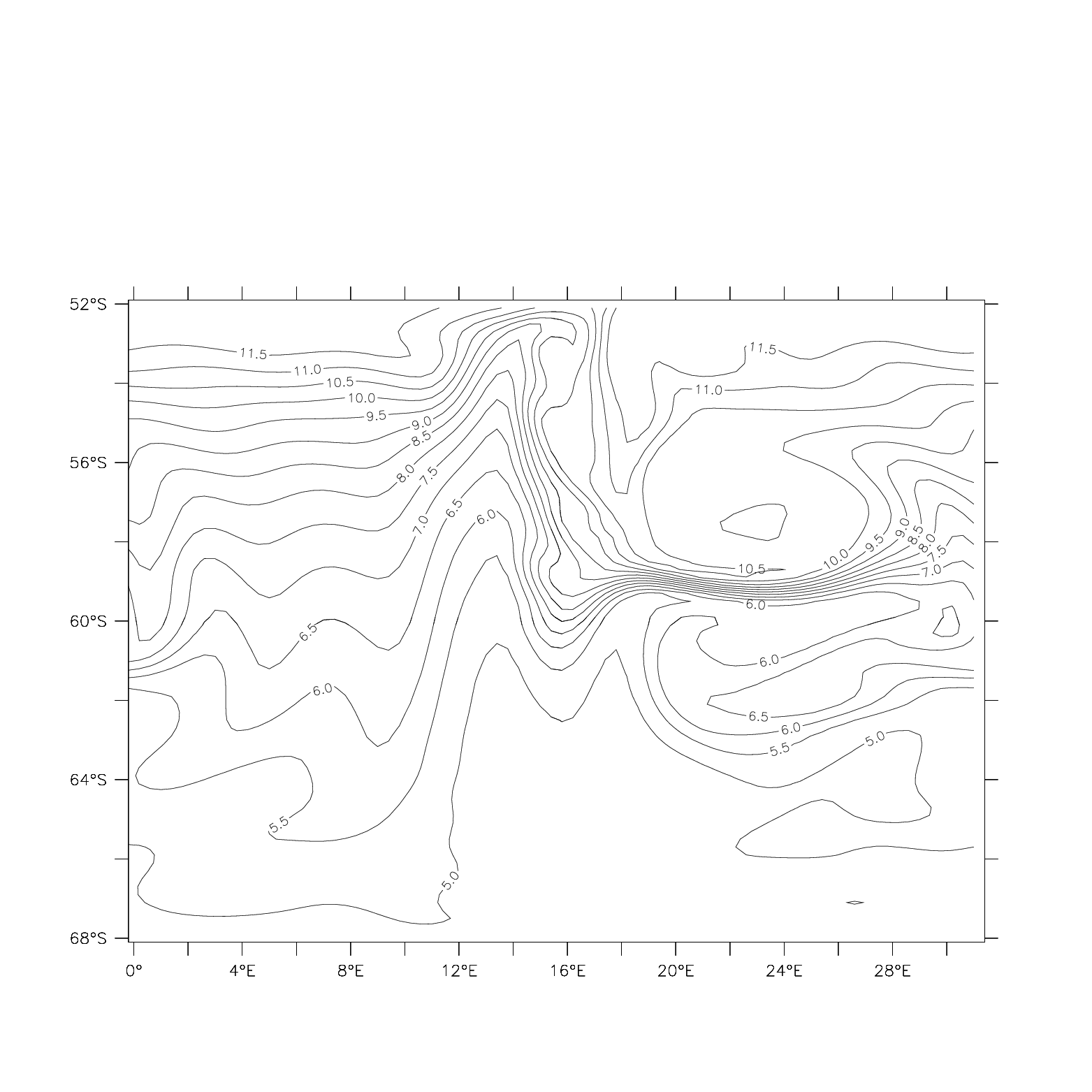}}\\
(a) Surface pot. temp., full algorithm &
(b) Surface pot. temp., reduced algorithm \\
\scalebox{.5}{\includegraphics{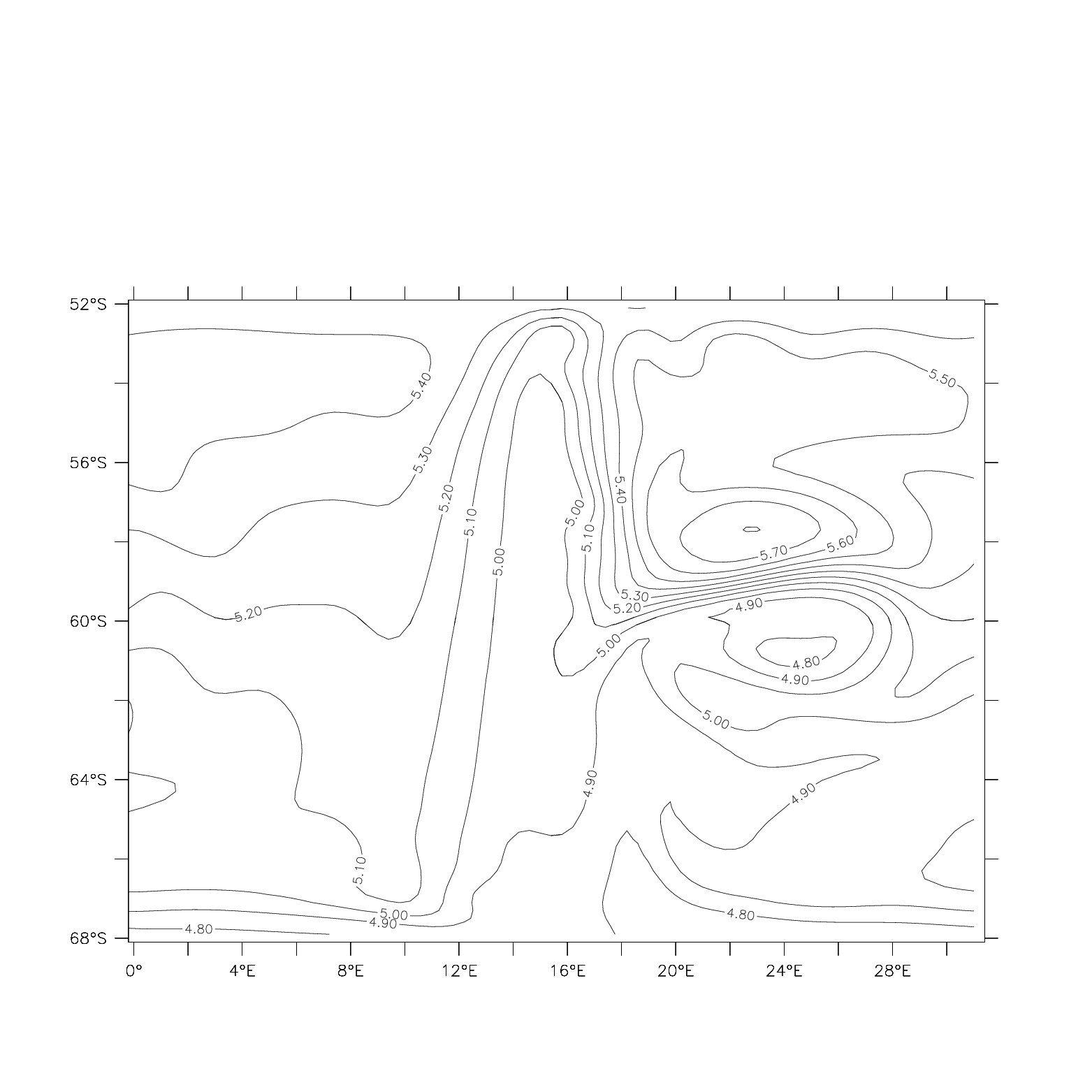}} &
\scalebox{.5}{\includegraphics{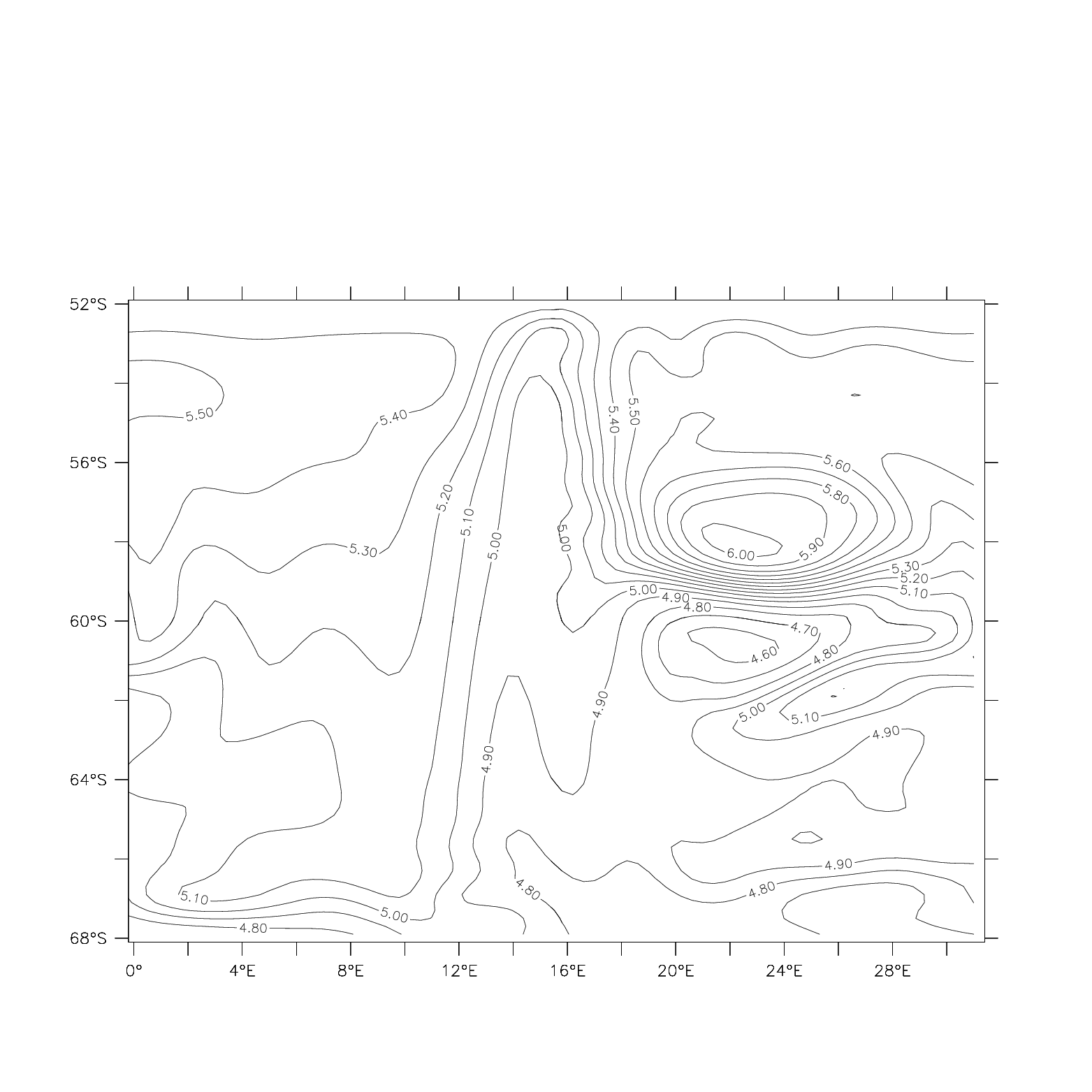}}\\
(c) Pot. temp. at 1600m, full algorithm &
(d) Pot. temp. at 1600m, reduced algorithm 
\end{tabular}
\caption{\label{f_T_sections} 
Snapshots of potential temperature in $^o$C at 150 years for full and reduced algorithm for experiment 0.4F.  This shows that the two algorithms produce nearly identical temperature fields.  
}\end{figure}

The purpose of this section is to show that the full and reduced POP-$\alpha$ algorithms produce nearly identical results.  A qualitative comparison of temperature and velocity fields of simulation 0.4F after 150 years shows that the dynamics are essentially the same (Figs. \ref{f_vel_sections} and \ref{f_T_sections}).  Minor differences in the strength and location of eddies are due to the chaotic, time-varying nature of these structures.  A more comprehensive comparison involves quantitative global statistics such as potential temperature and kinetic energy averaged over the domain.  The difference between the full and reduced POP-$\alpha$ algorithms is less than 0.05\%  in potential temperature (Fig. \ref{f_T_KE_diff}a), and less than 1\% in kinetic energy for low resolution experiments (Fig. \ref{f_T_KE_diff}b).  Differences of up to 20\% in global kinetic energy for the 0.4 cases are due to the high variability caused by transient eddies (Fig. \ref{f_T_KE}b).

\begin{figure}[tbh]
\scalebox{1}{\includegraphics{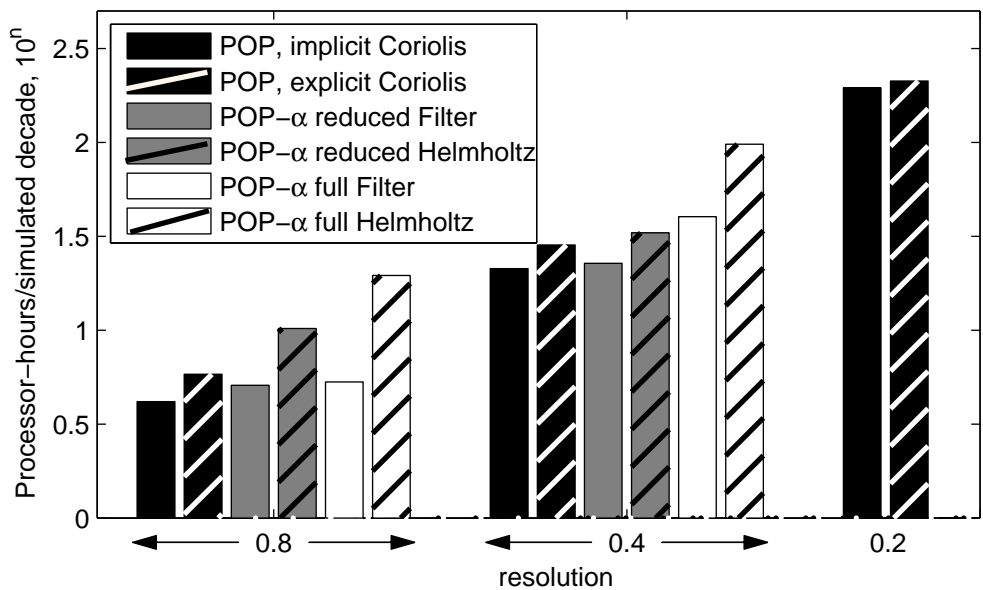}}
\caption{\label{f_timing} 
Timing data from various implementations of POP and POP-$\alpha$.  All simulations that use the Helmholtz inversion to smooth are extremely slow.  The POP-$\alpha$ algorithm can be sped up in two ways: using a filter, rather than a Helmholtz inversion to smooth; and changing from the full to the reduced algorithm.  All POP-$\alpha$ simulations use explicit Coriolis discretization.  Both implicit and explicit Coriolis simulations of POP are shown for comparison.
}\end{figure}

The general goal of turbulence modeling is to produce results
similar to higher resolution simulations without the computational
cost of the higher resolution.  Thus the running time of POP-$\alpha$
versus POP is a critical metric for the success of POP-$\alpha$.  The
full POP-$\alpha$ algorithm using a Helmholtz inversion is nearly as
costly as a doubling of resolution (Fig. \ref{f_timing}, Table
\ref{t_timing}); thus it is not a viable option for a turbulence
model.  The reduced algorithm is faster than the full algorithm.
Switching from a Helmholtz inversion, which requires a CG iterative
method for each smoothing step, to a simple filter that averages
nearest neighbors makes it faster still.  A full comparison of timing
and performance of various smoothing methods is detailed in
\cite{Petersen07jcp}.  The point here is that an efficient model
results from use of the reduced, rather than the full, algorithm, with
nearly identical results.

An important feature of the POP-$\alpha$ algorithm is that a longer
time-step can be taken with it than with standard POP (Table
\ref{t_timing}), consistent with the findings of
\cite{Wingate04mwr}. The reduced POP-$\alpha$ algorithm with a filter
is actually faster than POP due to this relaxed time step restriction,
so long as the comparison is restricted to cases making use of an
explicit treatment of the Coriolis term.  Even when POP is used with
the advantageous implicit treatment of the Coriolis term the best
POP-$\alpha$ algorithm only takes 6.5\% longer than the best POP time
(at 0.4 resolution).  This compares to a doubling of resolution of
standard POP that takes nine times as long.

The efficiency of the LANS-$\alpha$ model depends on the value of
$\alpha$ in the Helmholtz inversion, or equivalently the size of the
averaging stencil for the filter.  As $\alpha\rightarrow 0$ the
LANS-$\alpha$ equations return to the Navier-Stokes equations; larger
$\alpha$ makes the smooth velocity smoother, so that the LANS-$\alpha$
model has greater effect.  In the results presented here,
$\alpha=\Delta x$, the width of one grid-cell.  As $\alpha$ reaches a
value in the range of 2 to $2.5 \Delta x$ the global kinetic energy
grows and the simulation becomes unstable.  Analogously, simulations
may become unstable with larger filter widths.  This instability can
be countered somewhat with higher viscosity, but a threshold of
instability for large $\alpha$ still exists.

We found that the reduced algorithm is less sensitive to this
instability than the full algorithm.  For example, in the 0.4 case,
both full and reduced algorithms run stably for $\alpha=1.5\Delta x$;
when $\alpha=2 \Delta x$ the reduced algorithm is stable but the full
algorithm is not; when $\alpha=2.5 \Delta x$ both algorithms are
unstable.  The filter produces similar results, where the reduced
algorithm is stable for a wider range of filters than the full
algorithm.  This instability manifests itself in the barotropic
solver, where the iterative CG routine does not converge.  It is not
surprising that the full algorithm is more sensitive to instabilities
than the reduced algorithm in the barotropic solver; in the full
algorithm, each iteration of the CG solver includes a smoothing step
that requires a nested CG solver for the Helmholtz inversion.
Generally, this type of nested iteration is poor algorithm design.

\section{Conclusions \label{s_conclusions}}

Implementation of the LANS-$\alpha$ turbulence parameterization in a
primitive equation ocean model raises a number of new issues. The
expediency of substituting a local smoothing filter for the global
inverse Helmholtz operation has been previously established
\cite{Geurts_Holm05jt} but the appearance of the operator within the
barotropic mode equation raises an additional challenge to efficient
implementation.

We have presented here the details of a fundamental implementation of
LANS-$\alpha$ within a primitive equation ocean general circulation,
with that form referred to here as the full POP-$\alpha$
algorithm. Results from the full algorithm have been shown to be
consistent with those from previous geophysical studies of
LANS-$\alpha$ in simpler models
\cite{Holm_Nadiga03jpo,Wingate04mwr,Holm_Wingate05jpo}. With this full
algorithm as a point of reference we have found an alternative
implementation, which we refer to as the reduced POP-$\alpha$
algorithm and which produces nearly identical results; the step that
is skipped in the reduced form was anticipated to have little impact
due to the known tolerance for approximation in the barotropic set of
equations.  Either algorithm can be used with local filtering in
place of the global Helmholtz inversion.

The reduced form of the POP-$\alpha$ algorithm used in conjunction
with local filtering in place of the much more costly global Helmholtz
inversion produces a model which is only slightly more expensive than
plain POP on a per-time-step basis. The longer time step, which is
possible with inclusion of LANS-$\alpha$, is countered in our
experience by our inability to achieve a stable form of the algorithm
with an implicit form of the Coriolis term; even so the overall
increase in cost with use of our most efficient implementation is very
small in comparison with the cost associated with a doubling of model
resolution, making POP-$\alpha$ an attractive and potentially powerful
option for ocean climate modeling.

The linear stability analysis of the continuous equations and
algorithm gives some insight into how the LANS-alpha model improves
turbulence characteristics: the LANS-$\alpha$ model effectively makes
the Rossby radius of deformation larger.  Typical ocean-climate
simulations either don't resolve or just barely resolve the Rossby
radius; yet this scale is critical, as it is the size of ocean eddies
and is the scale where kinetic energy forcing occurs due to the
baroclinic instability.  Because of the way the LANS-$\alpha$ model
accounts for the effects of the small scale on the large, it generates
an 'effective' Rossby radius that is larger than in the
unaveraged case, making the effects of baroclinic instability
resolvable on coarser meshes than it could normally appear.  
One might object that a larger Rossby radius is an unphysical
representation of the original primitive equation set.  This would be
a valid objection if those scales were well resolved in the first
place.  However, in most ocean-climate simulations today the Rossby
radius is smaller than a grid-cell width or as large as a few
grid-cells.  Obviously, the turbulence and kinetic energy forcing
cannot be well-represented when it is so completely underresolved.  By
making the effective Rossby radius larger, the LANS-$\alpha$ model
circumvents this particular deficiency of low resolution, so that
global statistics related to baroclinic instability more closely
resemble those of higher resolution simulations.

We expect the LANS-$\alpha$ turbulence parameterization to be
particularly effective at an ocean model grid resolution that has been
coarsened by approximately a factor of two to four, relative to the
resolution required to bring out a vigorous mesoscale eddy
field. So-called eddy-resolving ocean modeling, capturing such a
vigorous eddy field, is known to require a grid resolution on the
order of $0.1^\circ$ \cite{Smith_ea00jpo}. Future work should address
the question of the interaction of LANS-$\alpha$ with
Gent-McWilliams-style isopycnal tracer mixing. If the two
parameterizations are found to be compatible, as would seem likely,
and if furthermore the oceanic jet systems such as the Gulf
Stream/North Atlantic Current come into much more realistic focus with
the inclusion of strong eddy transport of momentum at what would
otherwise be non-eddy-resolving scale, then we enter a new regime of
climate modeling with much more powerful modeling capabilities at our
disposal, enabling long simulations with a more detailed and more
faithful representation of the climate system. This statement is
predicated on not one but two caveats. There is more work yet to be
done, but we believe the results found thus far justify such work
towards an ambitious goal.

\section{Acknowledgements}
This work was carried out under the auspices of the National Nuclear
Security Administration of the U.S. Department of Energy at Los Alamos
National Laboratory under Contract No. DE-AC52-06NA25396.



\bibliographystyle{elsart-num}
\bibliography{alpha_model,ocean_modeling,dynamics,my_pubs}



\clearpage

\clearpage

\clearpage
\newcommand{\Dx}[1]{$\Delta x$}
\begin{table}
\center
  \begin{tabular}[c]{l|cccccccc}
   name & model & smoothing & $\alpha$ & fw &grid & lon & lat\\
   \hline
   \hline
   0.8 & POP & - & - & - &  40x40x34  & 0.8 & 0.4 \\
   0.4 & POP & - & - & - &  80x80x34  & 0.4 & 0.2 \\
   0.2 & POP & - & - & - &  160x160x34  & 0.2 & 0.1 \\
   \hline
   0.8H & POP-$\alpha$ & Helmholtz & 1.0\Dx &  - & 40x40x34  & 0.8 & 0.4\\ 
   0.4H & POP-$\alpha$ & Helmholtz & 1.0\Dx &  - & 80x80x34  & 0.4 & 0.2\\ 
   \hline
   0.8F & POP-$\alpha$ & filter & - & 3 & 40x40x34  & 0.8 & 0.4\\ 
   0.4F & POP-$\alpha$ & filter & - & 3 & 80x80x34  & 0.4 & 0.2\\ 
 \end{tabular}
\caption{\label{t_parameters}
Model parameters for experiments discussed in this paper, where {\it fw} is the filter width; {\it grid} is the number of gridpoints in $(x,y,z)$;  {\it lon} is the longitudinal grid-cell width; and {\it lat} is the latitudinal grid-cell width.  The names correspond to the meridional resolution and type of smoothing.  All POP-$\alpha$ simulations were run with both the full and reduced algorithms.
} \end{table}

\begin{table}
\center
  \begin{tabular}[c]{cc|ccc|r@{.}lcc}
{\bf algorithm} & {\bf Cor} & \multicolumn{3}{c}{\bf steps/day} & \multicolumn{4}{c}{\bf clock time} \\
\hline
 & & \multicolumn{3}{c}{resolution} & \multicolumn{4}{c}{resolution} \\
 & &   0.8 & 0.4 & 0.2 &   0&8 & 0.4 & 0.2 \\
   \hline 
POP & imp & 12 & 22 & 40 & 4&16 & 21.3 & 195\\
POP & exp & 20 & 32 & 52 & 5&82 & 28.4 & 212\\
red. POP-$\alpha$ filter & exp & 18 & 24 &  & 5&09 & 22.7 & \\
red. POP-$\alpha$ Helm.  & exp & 12 & 18 &  &10&21 & 33.0 & \\
full POP-$\alpha$ filter & exp & 16 & 24 &  & 5&30 & 40.2 & \\
full POP-$\alpha$ Helm.  & exp & 12 & 18 &  &19&56 & 97.8 & \\
   \hline
 \end{tabular}
\caption{\label{t_timing}
Minimimum steps/day and the resulting clock time for various algorithms and resoltions.  The Cor column states whether the barotropic Coriolis term is implicit or exlpicit.  Clock time is in processor-hours per simulated decade.  The fastest POP-$\alpha$ algorithm is the reduced algorithm with a filter.  Even though the POP-$\alpha$ algorithms use explicit Coriolis discretization, the timestep is smaller than standard POP with explicit Coriolis.
} \end{table}

\end{document}